\documentclass[aps,prl,preprint,amsmath,amssymb,preprintnumbers,groupedaddress]{revtex4-1}
\usepackage{CJK}
\usepackage{graphicx}
\usepackage{dcolumn}
\usepackage{bm}
\usepackage{color}
\usepackage{hyperref}

\begin{document}


\title{Dynamics of the linear-chain alpha cluster in microscopic time-dependent relativistic density functional theory}

\author{Z. X. Ren}
\affiliation{State Key Laboratory of Nuclear Physics and Technology, School of Physics, Peking University, Beijing 100871, China}

\author{P. W. Zhao}
\email{pwzhao@pku.edu.cn}
\affiliation{State Key Laboratory of Nuclear Physics and Technology, School of Physics, Peking University, Beijing 100871, China}

\author{J. Meng}
\email{mengj@pku.edu.cn}
\affiliation{State Key Laboratory of Nuclear Physics and Technology, School of Physics, Peking University, Beijing 100871, China}
\affiliation{School of Physics and Nuclear Energy Engineering, Beihang University, Beijing 100191, China}
\affiliation{Yukawa Institute for Theoretical Physics, Kyoto University, Kyoto 606-8502, Japan}

\begin{abstract}
  The time-dependent covariant density functional theory in 3D lattice space has been developed and applied to investigate the microscopic dynamics of the linear-chain cluster states for carbon isotopes in the reactions $^4$He$+^8$Be and $^4$He$+^{10}$Be without any symmetry assumptions.
  By examining the density distribution and its time evolutions, the structure and dynamics of the linear-chain states are analyzed, and the quasiperiodic oscillations of the clusters are revealed.
  For $^4$He$+^8$Be, the linear-chain states evolve to a triangular configuration and then to a more compact shape.
  In contrast, for $^4$He$+^{10}$Be, the lifetime of the linear-chain states is much more prolonged due to the dynamical isospin effects by the valence neutrons which slow down the longitudinal oscillations of the clusters and persist the linear-chain states.
  The dependence of the linear chain survival time and dynamical isospin effects on impact parameters have been illustrated as well.
\end{abstract}

\maketitle

Nuclear exotic deformations provide us an excellent framework to investigate the fundamental properties of quantum many-body systems~\cite{Wheeler_toroidal, Morinaga1956LinearChain, Wong1973Troidal, Cao2019Toroidal}.
For light nuclei, in particular, many exotic states may exist due to the appearance of the $\alpha$ cluster structure~\cite{freer2017microscopic}.
One of the most intriguing states among them is the linear-chain cluster states in carbon isotopes.

The linear-chain structure of three $\alpha$ clusters was suggested more than 60 years ago~\cite{Morinaga1956LinearChain}, and was used to explain the structure of the Hoyle state (the second $0^+$ state at $E_x = 7.65$ MeV in $^{12}$C), which plays a crucial role in the synthesis of $^{12}$C from three $^4$He nuclei in stars~\cite{hoyle1954nuclear}.
The Hoyle state was later found to be a gas-like state with strong mixing of the linear chain and other configurations~\cite{Fujiwara1980AlphaNuclei} and recently reinterpreted as an $\alpha$-condensate-like state~\cite{Tohsaki2001AlphaClusterC12_O16,Suhara2014ClusterCondensation}.
However, the concept of the linear-chain cluster state has attracted a lot of attentions for nuclear physicists, both experimentally and theoretically.
Its realization would have a strong impact on the research field of quantum many-body systems, because such an exotic clustering state is naturally recognized as an extreme of cluster structure due to its presumed propensity to exhibit bending configurations.

Various theoretical and experimental studies of linear-chain states have been carried out in $N=Z$ nuclei, such as $^{12}$C~\cite{Zhao2015Rod-shaped, Ren2019C12LCS, Liu2012CPC_C12O16Cluster, Horiuchi1975C12Cluster, Afanasjev2018C12_Cluster,}, $^{16}$O~\cite{Yao2014searching, Chevallier1967O16LCS, Ichikawa2011O16LinearChain}, $^{24}$Mg~\cite{Wuosmaa1992Mg24LCS, Rae1992Mg24, Iwata2015PRC_LCS_Mg, Inakura2018Rod_shaped}, etc. Further investigations are needed to confirm, however.
It is indeed very difficult to stabilize a linear-chain configuration against the bending motion because of the antisymmetrization effects and the weak-coupling nature of a clustering structure.
Nevertheless, Itagaki et al.,~\cite{Itagaki2001MolecularOrbit} suggested that a higher stability for the linear-chain states is possible in the neutron-rich nuclei.
This is because, if the valence neutrons occupy the so-called $\sigma$ orbital (orbital parallel to the symmetry axis), an elongated shape for the core would be favored to lower the energy of the valence neutrons.
The effects of the valence neutrons on cluster structure have been extensively investigated from both experimental~\cite{Freer2006Phys.Rev.Lett.42501,Navin2000Phys.Rev.Lett.266} and theoretical sides~\cite{Itagaki2000Phys.Rev.C44306, Oertzenn2004C14_LCS, Ito2008Phys.Rev.Lett.182502, Maruhn2010NPA_LCS_Carbon, Baba2014Phys.Rev.C64319}, and they are also found to be coherently coupled with the rotation of the whole system~\cite{Zhao2015Rod-shaped}.

The linear-chain states in carbon isotopes have been studied experimentally by the resonant scattering method with radioactive beams~\cite{Freer2011PRC13C, Freer2014C14_LCS_C14, Fritsch2016C14LCS, Yamaguchi2017C14LCS}.
In particular, the very recent measurement of the scattering $^4$He$+^{10}$Be has provided strong indications for the existence of linear-chain states in $^{14}$C~\cite{Yamaguchi2017C14LCS}.
The level spacing and relative energies to the $^4$He$+^{10}$Be threshold of the observed states agree well with the predictions of the antisymmetrized molecular dynamics method~\cite{Suhara2010C14ClusterExcited}, and the corresponding moment of inertia is also very close to the microscopic covariant density functional theory predictions~\cite{Zhao2015Rod-shaped}.

In despite of these accomplishments, until now there has been no theoretical discussion on the dynamics of the linear-chain states from the resonant scattering $^4$He$+^{10}$Be.
It is generally acknowledged that the time-dependent density functional theory (DFT) provides a powerful and versatile tool for simulating a great variety of dynamical scenarios~\cite{Negele1982TDDFT, Simene2012PEPJA, NakatsukasaRMP2016, SIMENEL2018TDHF_PPNP, STEVENSON2019PPNP}.
Therefore, it is interesting to investigate the formation and stability of the linear-chain configuration using the fully microscopic and dynamical time-dependent DFTs.
In fact, the three $\alpha$ linear-chain configuration of $^{12}$C has been studied by simulating the reaction $^4$He$+^8$Be with the time-dependent Hartree-Fock (TDHF) approach in the framework of the nonrelativistic DFTs~\cite{Umar2010TDHF_C12}.

The time-dependent mean field approach is also available in the framework of covariant DFTs, which bring many advantages in describing nuclear systems, such as the natural inclusion of the spin-orbit potential~\cite{RING1996PPNP, Vretenar2005PhysicsReport, meng2006PPNP}, and the self-consistent treatment of the time-odd fields~\cite{Meng2013FT_TAC}, see also Ref.~\cite{meng2016relativistic} for details.
However, due to the lack of computational resources, these calculations are limited with the simplified effective interactions~\cite{Cusson1985TDCDFT, Bai1987TDCDFT} or unphysical symmetry assumptions, such as the oscillations being restricted to axial symmetry~\cite{Vretenar1993TDRMF, VRETENAR1995TDRMF}.
Recently, the covariant DFT has been solved successfully in a three-dimensional (3D) lattice space with the inverse Hamiltonian~\cite{tanimura20153d} and Fourier spectral methods~\cite{REN2017Dirac3D, Ren2019C12LCS}.
This paves the way to develop the corresponding time-dependent approaches in a full 3D lattice space without assuming any symmetries.

In this Letter, the time-dependent covariant density functional theory (TDCDFT) is developed in a full 3D lattice space with modern relativistic density functionals.
Moreover, since microscopic dynamical studies of the linear-chain states in carbon isotopes with excess neutrons are still missing and strongly desired,
the newly developed approach is applied for a systematic investigation for the dynamics of the linear-chain states in $^{12}$C and $^{14}$C from respectively the resonant scatterings $^4$He$+^{8}$Be and $^4$He$+^{10}$Be.

Similar to the static CDFT, the starting point of TDCDFT is a standard effective Lagrangian density, where the nucleons can be coupled with either finite-range meson fields~\cite{Long2004PK1, lalazissis2005new} or zero-range point-coupling interactions~\cite{Niksic2008DDPC1, ZhaoPC-PK1}.
In the finite-range scheme, the exchanges of $\sigma$, $\omega$, and $\rho$ mesons between nucleons are usually considered.
Similarly, in the point-coupling scheme, the exchange of the various mesons is replaced by the corresponding contact interactions in the scalar-isoscalar, vector-isoscalar, and vector-isovector channels. The detailed formalism of the Lagrangian can be seen, e.g., in Ref.~\cite{meng2016relativistic}.

In the framework of the TDCDFT, one can derive the equation of motion for nucleons from the Lagrangian via a standard variational procedure~\cite{Vretenar2005PhysicsReport},
\begin{align}\label{eq_TDDIRAC}
   &i\partial_t\psi_k=[\bm{\alpha\cdot}(-i\bm{\nabla}-\bm{V})+V^0+\beta(M+S)]\psi_k,
\end{align}
which has the form of a time-dependent Dirac equation for nucleons with the mass $M$.
Here, $S$ and $V^\mu$ are the large scalar and vector potentials, respectively, and they are coupled with the corresponding densities and currents in a self-consistent way.
The densities and currents in the scalar, vector and isovector channels are time-dependent and are obtained from the single-particle wavefunctions $\psi_k$,
\begin{subequations}
  \begin{align}
    &\rho_S(\bm{r},t)=\sum_{k=1}^A\bar{\psi}_k\psi_k,\\
    &j_V^\mu(\bm{r},t)=\sum_{k=1}^A\bar{\psi}_k\gamma^\mu\psi_k,\\
    &j_{TV}^\mu(\bm{r},t)=\sum_{k=1}^A\bar{\psi}_k\tau_3\gamma^\mu\psi_k,
  \end{align}
\end{subequations}
where $\tau_3$ is the isospin Pauli matrix with the eigenvalues $+1$ for neutron and $-1$ for proton.

For the collisions of two nuclei, the initial wavefunctions of $\psi_k$ in Eq. \eqref{eq_TDDIRAC} are composed of the single-particle wavefunctions of the two nuclei, which are usually in their ground states, and are obtained from two separate static CDFT calculations.
Subsequently, the two nuclei are placed on the mesh with a large enough distance between them, so that the overlap between the wavefunctions of the two nuclei is negligible at the initial time.
Moreover, to set them in motion, each nucleus is Lorentz boosted to the given initial energy in the center-of-mass frame.
Similar to the TDHF calculations, an absorbing imaginary potential $iW(\bm{r})$~\cite{NakatsukasaRMP2016, SIMENEL2018TDHF_PPNP} is introduced to the vector potential $V^0(\bm{r},t)$ in Eq. \eqref{eq_TDDIRAC}, to remove the effects of the emitted particles.

The density functional DD-ME2~\cite{lalazissis2005new} is employed in the present calculations.
Using other density functional, such as PC-F1~\cite{Burvenich2002PCF1}, similar results can be obtained, and all the physical conclusions obtained based on the functional DD-ME2 remain unchanged.
The Dirac equation \eqref{eq_TDDIRAC} is solved in the real spacetime.
The Dirac spinors of the nucleons and the potentials are represented in 3D lattice space without any symmetry restriction.
The step sizes along the $x$, $y$, and $z$ axes are identical and chosen as 0.8 fm.
The predictor-corrector strategy~\cite{Simene2012PEPJA, Maruhn2014CPC} is adopted to solve the Dirac equation \eqref{eq_TDDIRAC} by expanding the time-evolution operator up to the fourth order.
The time step is set to be $\Delta t=0.1$ fm/$c$, and the total evolution time is chosen as 5000 fm/$c$.
The Poisson equation for the Coulomb potential is solved by the Hockney's method with the isolated boundary condition~\cite{eastwood1979remarks}.

The ground states of $^4$He, $^8$Be, and $^{10}$Be are calculated in a box with $24\times24\times24$ grid points, while for two nuclei collisions, a box with $30\times30\times50$ grid points is used.
The convergence with respect to the size of the box has been examined.
The calculated ground states of $^8$Be and $^{10}$Be are axially symmetric, and the symmetry axis is set to be the $z$ axis.
For the collisions of two nuclei, the initial velocities of the two nuclei are along the $z$ axis, and the initial separation distance
between the centers of the two nuclei is set to be 12 fm along $z$ axis.

\begin{figure}[!htbp]
  \centering
  \includegraphics[width=0.4\textwidth]{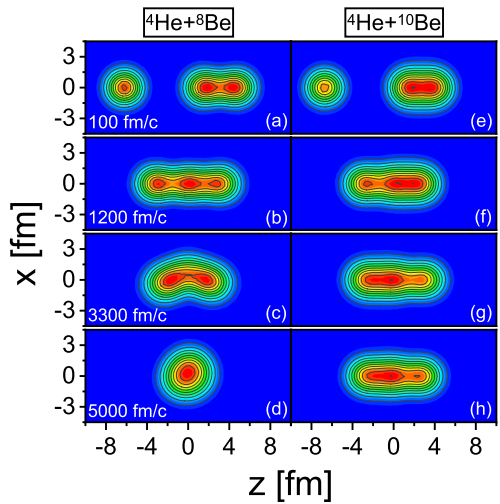}\\
  \caption{(color online). Selected total density distributions in the $y=0$ plane at the times $t=100$, $1200$, $3300$ and $5000$ fm/$c$ from TDCDFT time evolutions of the $^4$He$+^8$Be (left) and $^4$He$+^{10}$Be (right) head-on collisions. The initial energy is $E_{\rm c.m.}=2$~MeV.
  }\label{fig1}
\end{figure}

We have studied two collisions $^4$He$+^8$Be and $^4$He$+^{10}$Be, which lead to the linear-chain configurations in $^{12}$C and $^{14}$C, respectively.
Similar to Ref.~\cite{Umar2010TDHF_C12}, the initial energy is chosen as $E_{\rm c.m.}=2$ MeV to assure the occurrence of fusions.
In Fig.~\ref{fig1}, four snapshots from the long time evolution of the two collisions at zero impact parameters are presented in terms of the total density distributions at the times $t=100$, $1200$, $3300$ and $5000$ fm/$c$.
It is clear that the linear-chain states are formed in both collisions [see Figs.~\ref{fig1}(b) and (f)], but there are also distinct features.

For the $^4$He$+^8$Be collision, the system retains the linear-chain configuration up to the time at about 3100 fm/$c$, and then it starts to bend and forms a somewhat triangular-like shape as shown in Fig.~\ref{fig1}(c).
After staying at the triangular configuration for approximately 1000 fm/$c$, finally, the system relaxes into a more compact shape at even longer times [see Fig.~\ref{fig1}(d)].
Similar to Ref.~\cite{Umar2010TDHF_C12}, the head-on collision with the impact parameter $b=0$ here should be interpreted as the average value $\langle b\rangle=0$ with a small dispersion around this value.
This induces the breaking of the axial symmetry, and allows the bending motion and even the collapse to a more compact configuration.
It should be mentioned that during the time evolution of the linear-chain configuration, the three $\alpha$-like clusters are moving and do not equilibrate, but they exhibit a complex quasiperiodic oscillating motion with a little damping.
These features are indeed similar to the ones obtained by nonrelativistic TDHF calculations in Ref.~\cite{Umar2010TDHF_C12}.

For the $^4$He$+^{10}$Be collision, however, the linear-chain configuration of the system can be retained even at the time up to 5000 fm/$c$.
Moreover, comparing Figs.~\ref{fig1} (f) and (g), one can see that the positions of the $\alpha$-like and $^{10}$Be-like clusters could be exchanged during the time evolution.
This indeed reflects the quasiperiodic oscillating motion of the $\alpha$-like and $^{10}$Be-like clusters in this collision.

\begin{figure}[!htbp]
  \centering
  \includegraphics[width=0.4\textwidth]{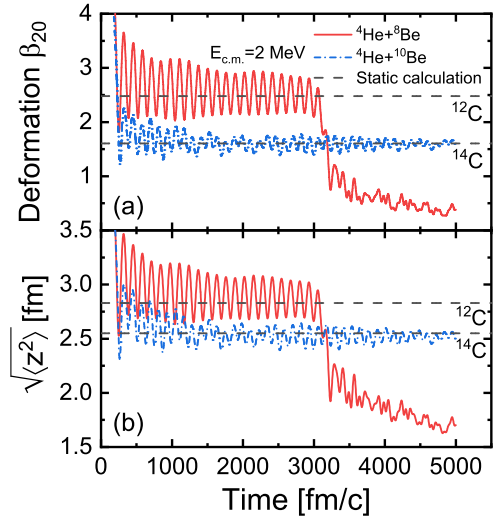}\\
  \caption{(color online). Time evolution of the quadrupole deformation $\beta_{20}$ (upper panel) and the length along $z$ direction $\sqrt{\langle z^2\rangle}$ (lower panel) for the head-on collisions of $^4$He$+^8$Be and $^4$He$+^{10}$Be systems at $E_{\rm c.m.}=2$~MeV.
  The results of the static CDFT calculations~\cite{Zhao2015Rod-shaped} are shown as the horizontal dashed lines.
  }\label{fig2}
\end{figure}

To investigate the dynamic quasiperiodic motion of the clusters in detail, in Fig.~\ref{fig2}(a), the time evolution of the quadrupole deformation $\beta_{20}$ for the head-on collisions of $^4$He$+^8$Be and $^4$He$+^{10}$Be systems are depicted.
For the $^4$He$+^8$Be collision, the $\beta_{20}$ oscillates in a quite harmonic way up to the time around 3100 fm/$c$, which corresponds to the linear-chain configuration.
Subsequently, a sudden decrease of $\beta_{20}$ appears and it corresponds to the occurrence of the bending [see Fig.~\ref{fig1} (c)], and then the $\beta_{20}$ gradually declines to values below 0.5 corresponding to a more compact shape [see Fig.~\ref{fig1} (d)].
With two more neutrons, for the $^4$He$+^{10}$Be collision, the $\beta_{20}$ also starts with a nearly harmonic oscillation, but up to a much longer time 5000 fm/$c$, which is the end of the time evolution in the present calculations.
This means that the linear-chain configuration could persist much longer in $^{14}$C than that in $^{12}$C.
Note that the amplitude of the $\beta_{20}$ oscillation is suppressed dramatically in the $^4$He$+^{10}$Be collision.
This may indicate that the two valence neutrons induce more attractions between the $\alpha$-like clusters and, thus, enhance the stability of the linear chain against the breathinglike breakup.

It is interesting to note that the $\beta_{20}$ obtained in static CDFT calculations~\cite{Zhao2015Rod-shaped} for the linear-chain states in $^{12}$C and $^{14}$C are also shown for comparison in Fig.~\ref{fig2}(a).
The results agree well with the present time-dependent calculations.
This means that the linear-chain configurations formed in the $^4$He$+^8$Be and $^4$He$+^{10}$Be collisions are consistent with the ones obtained in the nuclei $^{12}$C and $^{14}$C.

The oscillations of the $\beta_{20}$ can be explained by the time evolution of the length along the $z$ direction $\sqrt{\langle{z^2}\rangle}$ as shown in Fig.~\ref{fig2} (b).
For both $^4$He$+^8$Be and $^4$He$+^{10}$Be collisions, it is seen that the time evolution of $\sqrt{\langle{z^2}\rangle}$ exhibits very similar patterns to the evolution of the $\beta_{20}$.
This means that the oscillations of the $\beta_{20}$ are mainly from the cluster motions in the longitudinal direction.

\begin{figure}[!htbp]
  \centering
  \includegraphics[width=0.4\textwidth]{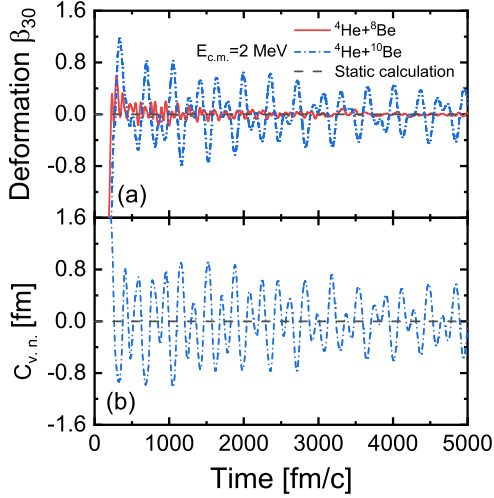}\\
  \caption{(color online). Same as Fig.~\ref{fig2} but for the octupole deformation $\beta_{30}$ (upper panel) and the center of valence neutrons in $z$ direction $C_{\rm v.n.}$ (see text) (lower panel).
  }\label{fig3}
\end{figure}

Apart from the quadrupole deformation $\beta_{20}$, it is also interesting to investigate the time evolution of the octupole deformation $\beta_{30}$, which reflects the asymmetry of the system with respect to the $x$-$y$ plane, and is shown in Fig.~\ref{fig3}(a) for the head-on collisions of $^4$He$+^8$Be and $^4$He$+^{10}$Be systems.
It is seen that the $\beta_{30}$ for both systems exhibits complex oscillations around $\beta_{30}=0$, in particular, when the systems are in the linear-chain configurations.
For the $^4$He$+^8$Be collision, the $\beta_{30}$ oscillation has a relatively higher frequency and a smaller amplitude, but it is opposite for the $^4$He$+^{10}$Be collision, where the frequency is lower and the amplitude is larger.
This reveals that the reflection symmetry with respect to the $x$-$y$ plane is not strongly broken during the $^4$He$+^8$Be collision, however, it is remarkably violated in the $^4$He$+^{10}$Be collision [see Figs.~\ref{fig1}(f)--(h)].

The distinct features of the $\beta_{30}$ oscillation in the $^4$He$+^{10}$Be collision are caused by the effects of two valence neutrons.
This can be seen in Fig.~\ref{fig3}(b), where the center of valence neutrons in the $z$ direction $C_{\rm v.n.}$ is defined,
\begin{equation}
   C_{\rm v.n.}=\frac{\int d^3{\bm r}~z(\rho_n-\rho_p)}{\int d^3{\bm r}~(\rho_n-\rho_p)},
\end{equation}
and its time evolution is depicted.
One can see that the $C_{\rm v.n.}$ oscillates in a very similar frequency with the one of the $\beta_{30}$.
This indicates that the two valence neutrons oscillate in the $z$ direction and, thus, induce the strong oscillation of the octupole deformation $\beta_{30}$.
There are discrepancies in the oscillating amplitudes of $C_{\rm v.n.}$ and $\beta_{30}$.
This is because that the magnitudes of the octupole deformation of the system, apart from the contribution of the valence neutrons, are also influenced by the asymmetric spatial distribution of the three $\alpha$s, which is mainly induced by the polarization effects for valence neutrons.

\begin{figure}[!htbp]
  \centering
  \includegraphics[width=0.4\textwidth]{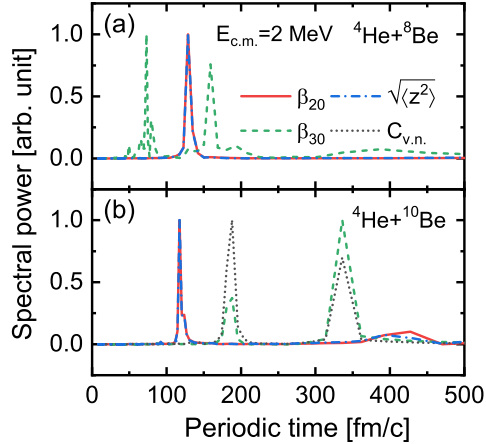}\\
  \caption{(color online). The absolute squares of the Fourier transformation (spectral power) of the quadrupole deformation $\beta_{20}$, the length along $z$ direction $\sqrt{\langle z^2\rangle}$, the octupole deformation $\beta_{30}$ (upper panel), and the center of valence neutrons in $z$ direction $C_{\rm v.n.}$ as a function of periodic time for the head-on collisions of $^4$He$+^8$Be (upper panel) and $^4$He$+^{10}$Be (lower panel) systems at $E_{\rm c.m.}=2$~MeV.
  }\label{fig4}
\end{figure}

For a quantitative study of the time evolutions of $\beta_{20}$, $\sqrt{\langle{z^2}\rangle}$, $\beta_{30}$, and $C_{\rm v.n.}$ for the linear-chain configurations formed in the head-on collisions of $^4$He$+^8$Be and $^4$He$+^{10}$Be systems, the Fourier transformations have been performed for the time surviving the linear-chain configurations.
In Fig.~\ref{fig4}, the obtained spectral powers for the $\beta_{20}$, $\sqrt{\langle{z^2}\rangle}$, $\beta_{30}$, and $C_{\rm v.n.}$ are shown as a function of the period time, which is just the inverse of the frequency.
By increasing the initial energy $E_{\rm c.m.}$ to 8~MeV, the amplitudes of oscillation are enhanced and the period times of oscillation become slightly longer, but no more than 10\%.
For both collisions, the spectral powers of $\beta_{20}$ are coincident with the ones of $\sqrt{\langle{z^2}\rangle}$, in particular for the positions of the peaks.
This reveals quantitatively that the oscillations of $\beta_{20}$ are mainly along the longitudinal directions in both collisions.

For the $^4$He$+^{10}$Be collision, it is interesting to see that the spectral powers of $\beta_{30}$ and $C_{\rm v.n.}$ are nicely matched, especially for the positions of the two peaks.
Note that the double-peak structure also appears in the $^4$He$+^{8}$Be collision, but at a higher frequency region [see Fig.~\ref{fig4}(a)].
Therefore, one can conclude that the two valence neutrons in the $^4$He$+^{10}$Be collision bring the dynamical isospin effects, which can slow down the $\beta_{30}$ oscillation in the $z$ direction and, thus, are helpful to persist the linear-chain configurations.

Quantitively, the period time of the oscillation of the valence neutrons is apparently different from the one of the three $\alpha$ clusters [represented by the $\beta_{20}$ oscillation in Fig.~\ref{fig4}(b)].
This reveals the weak coupling nature of the two valence neutrons and the three $\alpha$ clusters as the two valence neutrons occupy the $\pi$ orbits with density distribution perpendicular to that of the three $\alpha$ clusters.

In Ref.~\cite{Maruhn2010NPA_LCS_Carbon}, the stability of the linear-chain configurations of the neutron-rich $^{16}$C and $^{20}$C has been investigated based on the nonrelativistic TDHF approach.
It was found that the $\sigma$ orbit plays an important role for the increase of linear-chain lifetime.
Here for $^{14}$C, it is found that the $\pi$ orbit is also helpful to persist the linear-chain configurations due to the dynamical isospin effects.
In the future, it would be interesting to investigate the systems with $\sigma$ orbit by TDCDFT.

\begin{figure}[!htbp]
  \centering
  \includegraphics[width=0.4\textwidth]{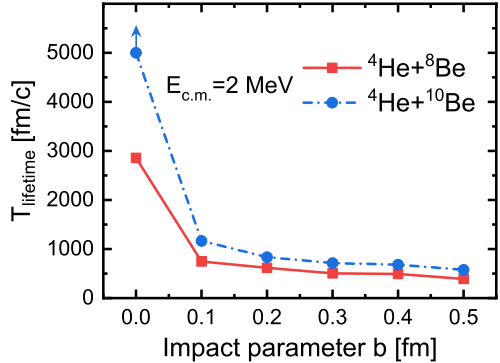}\\
  \caption{(color online). Lifetime of the linear-chain configuration as a function of the impact parameter $b$ for the $^4$He$+^8$Be and $^4$He$+^{10}$Be collisions at $E_{\rm c.m.}=2$~MeV.
  }\label{fig5}
\end{figure}

Finally, the stability of the linear-chain configurations with respect to different impact parameters $b$ has also been studied.
In Fig.~\ref{fig5}, the dependence of the linear chain survival time on the impact parameter for the $^4$He$+^8$Be and $^4$He$+^{10}$Be collisions is depicted.
It should be noted that, even taking the largest impact factor $b=0.5$ fm in Fig.~\ref{fig5}, the corresponding angular momentum is still around zero.
It is seen that as the impact parameter increases the survival time decreases rapidly, while at each impact parameter, the linear chain survival time for the $^4$He$+^{10}$Be system is longer than that for the $^4$He$+^8$Be system.
Moreover, it can be also seen that the linear-chain lifetime does not drop to zero but stays around 500 fm/$c$ with the increase of impact parameter.
Similar calculations have also been performed at center-of-mass energies $E_{\rm c.m.}=4$ and $8$ MeV, and it is found that the decreasing tendency of the linear-chain lifetime with the increasing impact parameter still holds true.

In summary, the time-dependent covariant density functional theory has been developed in full 3D lattice space without any symmetry assumptions, and it has been applied to investigate the microscopic dynamics of the linear-chain cluster states in $^{12}$C and $^{14}$C from respectively the resonant scatterings $^4$He$+^8$Be and $^4$He$+^{10}$Be.
The metastable linear-chain states are formed in both collisions, and the corresponding quadrupole deformations agree nicely with those obtained by the static calculations for $^{12}$C and $^{14}$C, respectively.
Quasiperiodic oscillations of the clusters are found in the time evolutions of densities for the linear-chain configurations.
For $^4$He$+^8$Be, the linear-chain states are followed by a transition to a lower-energy triangular configuration before acquiring a more compact shape.
In contrast, for $^4$He$+^{10}$Be, the linear-chain states could stay much longer due to the dynamical isospin effects of the two valence neutrons which slow down the longitudinal quasiperiodic oscillations of the clusters.
Finally, it is shown that with the increase of the impact parameter, the lifetime of the linear-chain states decreases rapidly, while the lifetime for the reaction $^4$He$+^{10}$Be is still longer than that for the reaction $^4$He$+^8$Be.

Adding neutrons is considered to be an important mechanism for the stability of the linear-chain configurations.
In the present work, we have provided the first investigation on the microscopic dynamics of the linear-chain structure with excess neutrons, and the dynamical isospin effects have been revealed.
With the development of worldwide facilities on rare isotope beams, exploring the linear-chain structure via resonant scatterings of rare isotope beams will be more and more
popular in experiment.
The present work has therefore also provided a promising theoretical tool to guide and interpret the future experiments.

\begin{acknowledgments}
This work was partly supported by the National Key R\&D Program of China (Contracts No. 2018YFA0404400 and No. 2017YFE0116700) and
the National Natural Science Foundation of China (Grants No. 11621131001, No. 11875075, No. 11935003, and No. 11975031).
\end{acknowledgments}


\begin{thebibliography}{60}%
\makeatletter
\providecommand \@ifxundefined [1]{%
 \@ifx{#1\undefined}
}%
\providecommand \@ifnum [1]{%
 \ifnum #1\expandafter \@firstoftwo
 \else \expandafter \@secondoftwo
 \fi
}%
\providecommand \@ifx [1]{%
 \ifx #1\expandafter \@firstoftwo
 \else \expandafter \@secondoftwo
 \fi
}%
\providecommand \natexlab [1]{#1}%
\providecommand \enquote  [1]{``#1''}%
\providecommand \bibnamefont  [1]{#1}%
\providecommand \bibfnamefont [1]{#1}%
\providecommand \citenamefont [1]{#1}%
\providecommand \href@noop [0]{\@secondoftwo}%
\providecommand \href [0]{\begingroup \@sanitize@url \@href}%
\providecommand \@href[1]{\@@startlink{#1}\@@href}%
\providecommand \@@href[1]{\endgroup#1\@@endlink}%
\providecommand \@sanitize@url [0]{\catcode `\\12\catcode `\$12\catcode
  `\&12\catcode `\#12\catcode `\^12\catcode `\_12\catcode `\%12\relax}%
\providecommand \@@startlink[1]{}%
\providecommand \@@endlink[0]{}%
\providecommand \url  [0]{\begingroup\@sanitize@url \@url }%
\providecommand \@url [1]{\endgroup\@href {#1}{\urlprefix }}%
\providecommand \urlprefix  [0]{URL }%
\providecommand \Eprint [0]{\href }%
\providecommand \doibase [0]{http://dx.doi.org/}%
\providecommand \selectlanguage [0]{\@gobble}%
\providecommand \bibinfo  [0]{\@secondoftwo}%
\providecommand \bibfield  [0]{\@secondoftwo}%
\providecommand \translation [1]{[#1]}%
\providecommand \BibitemOpen [0]{}%
\providecommand \bibitemStop [0]{}%
\providecommand \bibitemNoStop [0]{.\EOS\space}%
\providecommand \EOS [0]{\spacefactor3000\relax}%
\providecommand \BibitemShut  [1]{\csname bibitem#1\endcsname}%
\let\auto@bib@innerbib\@empty
\bibitem [{Whe()}]{Wheeler_toroidal}%
  \BibitemOpen
  \href@noop {} { {\bibinfo {title} {J. A. Wheeler, Nucleonics Notebook
  (unpublished, 1950); see also p. 297 in G. Gamow, Biography of Physics
  (Harper \& Brothers Publishers, New York, 1961).}}}\BibitemShut {Stop}%
\bibitem [{\citenamefont {Morinaga}(1956)}]{Morinaga1956LinearChain}%
  \BibitemOpen
  \bibfield  {author} {\bibinfo {author} {\bibfnamefont {H.}~\bibnamefont
  {Morinaga}},\ }\href {\doibase 10.1103/PhysRev.101.254} {\bibfield  {journal}
  {\bibinfo  {journal} {Phys. Rev.}\ }\textbf {\bibinfo {volume} {101}},\
  \bibinfo {pages} {254} (\bibinfo {year} {1956})}\BibitemShut {NoStop}%
\bibitem [{\citenamefont {Wong}(1973)}]{Wong1973Troidal}%
  \BibitemOpen
  \bibfield  {author} {\bibinfo {author} {\bibfnamefont {C.~Y.}\ \bibnamefont
  {Wong}},\ }\href {\doibase https://doi.org/10.1016/0003-4916(73)90420-X}
  {\bibfield  {journal} {\bibinfo  {journal} {Ann. Phys. (NY)}\ }\textbf
  {\bibinfo {volume} {77}},\ \bibinfo {pages} {279} (\bibinfo {year}
  {1973})}\BibitemShut {NoStop}%
\bibitem [{\citenamefont {Cao}\ \emph {et~al.}(2019)\citenamefont {Cao},
  \citenamefont {Kim}, \citenamefont {Schmidt}, \citenamefont {Hagel},
  \citenamefont {Barbui}, \citenamefont {Gauthier}, \citenamefont {Wuenschel},
  \citenamefont {Giuliani}, \citenamefont {Rodriguez}, \citenamefont
  {Kowalski}, \citenamefont {Zheng}, \citenamefont {Huang}, \citenamefont
  {Bonasera}, \citenamefont {Wada}, \citenamefont {Blando}, \citenamefont
  {Zhang}, \citenamefont {Wong}, \citenamefont {Staszczak}, \citenamefont
  {Ren}, \citenamefont {Wang}, \citenamefont {Zhang}, \citenamefont {Meng},\
  and\ \citenamefont {Natowitz}}]{Cao2019Toroidal}%
  \BibitemOpen
  \bibfield  {author} {\bibinfo {author} {\bibfnamefont {X.~G.}\ \bibnamefont
  {Cao}}, \bibinfo {author} {\bibfnamefont {E.~J.}\ \bibnamefont {Kim}},
  \bibinfo {author} {\bibfnamefont {K.}~\bibnamefont {Schmidt}}, \bibinfo
  {author} {\bibfnamefont {K.}~\bibnamefont {Hagel}}, \bibinfo {author}
  {\bibfnamefont {M.}~\bibnamefont {Barbui}}, \bibinfo {author} {\bibfnamefont
  {J.}~\bibnamefont {Gauthier}}, \bibinfo {author} {\bibfnamefont
  {S.}~\bibnamefont {Wuenschel}}, \bibinfo {author} {\bibfnamefont
  {G.}~\bibnamefont {Giuliani}}, \bibinfo {author} {\bibfnamefont {M.~R.~D.}\
  \bibnamefont {Rodriguez}}, \bibinfo {author} {\bibfnamefont {S.}~\bibnamefont
  {Kowalski}}, \bibinfo {author} {\bibfnamefont {H.}~\bibnamefont {Zheng}},
  \bibinfo {author} {\bibfnamefont {M.}~\bibnamefont {Huang}}, \bibinfo
  {author} {\bibfnamefont {A.}~\bibnamefont {Bonasera}}, \bibinfo {author}
  {\bibfnamefont {R.}~\bibnamefont {Wada}}, \bibinfo {author} {\bibfnamefont
  {N.}~\bibnamefont {Blando}}, \bibinfo {author} {\bibfnamefont {G.~Q.}\
  \bibnamefont {Zhang}}, \bibinfo {author} {\bibfnamefont {C.~Y.}\ \bibnamefont
  {Wong}}, \bibinfo {author} {\bibfnamefont {A.}~\bibnamefont {Staszczak}},
  \bibinfo {author} {\bibfnamefont {Z.~X.}\ \bibnamefont {Ren}}, \bibinfo
  {author} {\bibfnamefont {Y.~K.}\ \bibnamefont {Wang}}, \bibinfo {author}
  {\bibfnamefont {S.~Q.}\ \bibnamefont {Zhang}}, \bibinfo {author}
  {\bibfnamefont {J.}~\bibnamefont {Meng}}, \ and\ \bibinfo {author}
  {\bibfnamefont {J.~B.}\ \bibnamefont {Natowitz}},\ }\href {\doibase
  10.1103/PhysRevC.99.014606} {\bibfield  {journal} {\bibinfo  {journal} {Phys.
  Rev. C}\ }\textbf {\bibinfo {volume} {99}},\ \bibinfo {pages} {014606}
  (\bibinfo {year} {2019})}\BibitemShut {NoStop}%
\bibitem [{\citenamefont {Freer}\ \emph {et~al.}(2018)\citenamefont {Freer},
  \citenamefont {Horiuchi}, \citenamefont {Kanada-En'yo}, \citenamefont {Lee},\
  and\ \citenamefont {Mei\ss{}ner}}]{freer2017microscopic}%
  \BibitemOpen
  \bibfield  {author} {\bibinfo {author} {\bibfnamefont {M.}~\bibnamefont
  {Freer}}, \bibinfo {author} {\bibfnamefont {H.}~\bibnamefont {Horiuchi}},
  \bibinfo {author} {\bibfnamefont {Y.}~\bibnamefont {Kanada-En'yo}}, \bibinfo
  {author} {\bibfnamefont {D.}~\bibnamefont {Lee}}, \ and\ \bibinfo {author}
  {\bibfnamefont {U.-G.}\ \bibnamefont {Mei\ss{}ner}},\ }\href {\doibase
  10.1103/RevModPhys.90.035004} {\bibfield  {journal} {\bibinfo  {journal}
  {Rev. Mod. Phys.}\ }\textbf {\bibinfo {volume} {90}},\ \bibinfo {pages}
  {035004} (\bibinfo {year} {2018})}\BibitemShut {NoStop}%
\bibitem [{\citenamefont {Hoyle}(1954)}]{hoyle1954nuclear}%
  \BibitemOpen
  \bibfield  {author} {\bibinfo {author} {\bibfnamefont {F.}~\bibnamefont
  {Hoyle}},\ }\href@noop {} {\bibfield  {journal} {\bibinfo  {journal}
  {Astrophys. J. Suppl. Ser.}\ }\textbf {\bibinfo {volume} {1}},\ \bibinfo
  {pages} {121} (\bibinfo {year} {1954})}\BibitemShut {NoStop}%
\bibitem [{\citenamefont {Fujiwara}\ \emph {et~al.}(1980)\citenamefont
  {Fujiwara}, \citenamefont {Horiuchi}, \citenamefont {Ikeda}, \citenamefont
  {Kamimura}, \citenamefont {Kat\={o}}, \citenamefont {Suzuki},\ and\
  \citenamefont {Uegaki}}]{Fujiwara1980AlphaNuclei}%
  \BibitemOpen
  \bibfield  {author} {\bibinfo {author} {\bibfnamefont {Y.}~\bibnamefont
  {Fujiwara}}, \bibinfo {author} {\bibfnamefont {H.}~\bibnamefont {Horiuchi}},
  \bibinfo {author} {\bibfnamefont {K.}~\bibnamefont {Ikeda}}, \bibinfo
  {author} {\bibfnamefont {M.}~\bibnamefont {Kamimura}}, \bibinfo {author}
  {\bibfnamefont {K.}~\bibnamefont {Kat\={o}}}, \bibinfo {author}
  {\bibfnamefont {Y.}~\bibnamefont {Suzuki}}, \ and\ \bibinfo {author}
  {\bibfnamefont {E.}~\bibnamefont {Uegaki}},\ }\href {\doibase
  10.1143/PTPS.68.29} {\bibfield  {journal} {\bibinfo  {journal} {Prog. Theor.
  Phys.}\ }\textbf {\bibinfo {volume} {68}},\ \bibinfo {pages} {29} (\bibinfo
  {year} {1980})}\BibitemShut {NoStop}%
\bibitem [{\citenamefont {Tohsaki}\ \emph {et~al.}(2001)\citenamefont
  {Tohsaki}, \citenamefont {Horiuchi}, \citenamefont {Schuck},\ and\
  \citenamefont {R\"opke}}]{Tohsaki2001AlphaClusterC12_O16}%
  \BibitemOpen
  \bibfield  {author} {\bibinfo {author} {\bibfnamefont {A.}~\bibnamefont
  {Tohsaki}}, \bibinfo {author} {\bibfnamefont {H.}~\bibnamefont {Horiuchi}},
  \bibinfo {author} {\bibfnamefont {P.}~\bibnamefont {Schuck}}, \ and\ \bibinfo
  {author} {\bibfnamefont {G.}~\bibnamefont {R\"opke}},\ }\href {\doibase
  10.1103/PhysRevLett.87.192501} {\bibfield  {journal} {\bibinfo  {journal}
  {Phys. Rev. Lett.}\ }\textbf {\bibinfo {volume} {87}},\ \bibinfo {pages}
  {192501} (\bibinfo {year} {2001})}\BibitemShut {NoStop}%
\bibitem [{\citenamefont {Suhara}\ \emph {et~al.}(2014)\citenamefont {Suhara},
  \citenamefont {Funaki}, \citenamefont {Zhou}, \citenamefont {Horiuchi},\ and\
  \citenamefont {Tohsaki}}]{Suhara2014ClusterCondensation}%
  \BibitemOpen
  \bibfield  {author} {\bibinfo {author} {\bibfnamefont {T.}~\bibnamefont
  {Suhara}}, \bibinfo {author} {\bibfnamefont {Y.}~\bibnamefont {Funaki}},
  \bibinfo {author} {\bibfnamefont {B.}~\bibnamefont {Zhou}}, \bibinfo {author}
  {\bibfnamefont {H.}~\bibnamefont {Horiuchi}}, \ and\ \bibinfo {author}
  {\bibfnamefont {A.}~\bibnamefont {Tohsaki}},\ }\href {\doibase
  10.1103/PhysRevLett.112.062501} {\bibfield  {journal} {\bibinfo  {journal}
  {Phys. Rev. Lett.}\ }\textbf {\bibinfo {volume} {112}},\ \bibinfo {pages}
  {062501} (\bibinfo {year} {2014})}\BibitemShut {NoStop}%
\bibitem [{\citenamefont {Zhao}\ \emph {et~al.}(2015)\citenamefont {Zhao},
  \citenamefont {Itagaki},\ and\ \citenamefont {Meng}}]{Zhao2015Rod-shaped}%
  \BibitemOpen
  \bibfield  {author} {\bibinfo {author} {\bibfnamefont {P.~W.}\ \bibnamefont
  {Zhao}}, \bibinfo {author} {\bibfnamefont {N.}~\bibnamefont {Itagaki}}, \
  and\ \bibinfo {author} {\bibfnamefont {J.}~\bibnamefont {Meng}},\ }\href
  {\doibase 10.1103/PhysRevLett.115.022501} {\bibfield  {journal} {\bibinfo
  {journal} {Phys. Rev. Lett.}\ }\textbf {\bibinfo {volume} {115}},\ \bibinfo
  {pages} {022501} (\bibinfo {year} {2015})}\BibitemShut {NoStop}%
\bibitem [{\citenamefont {Ren}\ \emph {et~al.}(2019)\citenamefont {Ren},
  \citenamefont {Zhang}, \citenamefont {Zhao}, \citenamefont {Itagaki},
  \citenamefont {Maruhn},\ and\ \citenamefont {Meng}}]{Ren2019C12LCS}%
  \BibitemOpen
  \bibfield  {author} {\bibinfo {author} {\bibfnamefont {Z.~X.}\ \bibnamefont
  {Ren}}, \bibinfo {author} {\bibfnamefont {S.~Q.}\ \bibnamefont {Zhang}},
  \bibinfo {author} {\bibfnamefont {P.~W.}\ \bibnamefont {Zhao}}, \bibinfo
  {author} {\bibfnamefont {N.}~\bibnamefont {Itagaki}}, \bibinfo {author}
  {\bibfnamefont {J.~A.}\ \bibnamefont {Maruhn}}, \ and\ \bibinfo {author}
  {\bibfnamefont {J.}~\bibnamefont {Meng}},\ }\href {\doibase
  https://doi.org/10.1007/s11433-019-9412-3} {\bibfield  {journal} {\bibinfo
  {journal} {Sci. China-Phys. Mech. Astron.}\ }\textbf {\bibinfo {volume}
  {62}},\ \bibinfo {pages} {112062} (\bibinfo {year} {2019})}\BibitemShut
  {NoStop}%
\bibitem [{\citenamefont {Liu}\ and\ \citenamefont
  {Zhao}(2012)}]{Liu2012CPC_C12O16Cluster}%
  \BibitemOpen
  \bibfield  {author} {\bibinfo {author} {\bibfnamefont {L.}~\bibnamefont
  {Liu}}\ and\ \bibinfo {author} {\bibfnamefont {P.-W.}\ \bibnamefont {Zhao}},\
  }\href {http://stacks.iop.org/1674-1137/36/i=9/a=004} {\bibfield  {journal}
  {\bibinfo  {journal} {Chin. Phys. C}\ }\textbf {\bibinfo {volume} {36}},\
  \bibinfo {pages} {818} (\bibinfo {year} {2012})}\BibitemShut {NoStop}%
\bibitem [{\citenamefont {Horiuchi}(1975)}]{Horiuchi1975C12Cluster}%
  \BibitemOpen
  \bibfield  {author} {\bibinfo {author} {\bibfnamefont {H.}~\bibnamefont
  {Horiuchi}},\ }\href {\doibase 10.1143/PTP.53.447} {\bibfield  {journal}
  {\bibinfo  {journal} {Progress of Theoretical Physics}\ }\textbf {\bibinfo
  {volume} {53}},\ \bibinfo {pages} {447} (\bibinfo {year} {1975})}\BibitemShut
  {NoStop}%
\bibitem [{\citenamefont {Afanasjev}\ and\ \citenamefont
  {Abusara}(2018)}]{Afanasjev2018C12_Cluster}%
  \BibitemOpen
  \bibfield  {author} {\bibinfo {author} {\bibfnamefont {A.~V.}\ \bibnamefont
  {Afanasjev}}\ and\ \bibinfo {author} {\bibfnamefont {H.}~\bibnamefont
  {Abusara}},\ }\href {\doibase 10.1103/PhysRevC.97.024329} {\bibfield
  {journal} {\bibinfo  {journal} {Phys. Rev. C}\ }\textbf {\bibinfo {volume}
  {97}},\ \bibinfo {pages} {024329} (\bibinfo {year} {2018})}\BibitemShut
  {NoStop}%
\bibitem [{\citenamefont {Maruhn}\ \emph {et~al.}(2005)\citenamefont {Maruhn},
  \citenamefont {Reinhard}, \citenamefont {Stevenson}, \citenamefont {Stone},\
  and\ \citenamefont {Strayer}}]{}%
  \BibitemOpen
  \bibfield  {author} {\bibinfo {author} {\bibfnamefont {J.~A.}\ \bibnamefont
  {Maruhn}}, \bibinfo {author} {\bibfnamefont {P.~G.}\ \bibnamefont
  {Reinhard}}, \bibinfo {author} {\bibfnamefont {P.~D.}\ \bibnamefont
  {Stevenson}}, \bibinfo {author} {\bibfnamefont {J.~R.}\ \bibnamefont
  {Stone}}, \ and\ \bibinfo {author} {\bibfnamefont {M.~R.}\ \bibnamefont
  {Strayer}},\ }\href {\doibase 10.1103/PhysRevC.71.064328} {\bibfield
  {journal} {\bibinfo  {journal} {Phys. Rev. C}\ }\textbf {\bibinfo {volume}
  {71}},\ \bibinfo {pages} {064328} (\bibinfo {year} {2005})}\BibitemShut
  {NoStop}%
\bibitem [{\citenamefont {Yao}\ \emph {et~al.}(2014)\citenamefont {Yao},
  \citenamefont {Itagaki},\ and\ \citenamefont {Meng}}]{Yao2014searching}%
  \BibitemOpen
  \bibfield  {author} {\bibinfo {author} {\bibfnamefont {J.~M.}\ \bibnamefont
  {Yao}}, \bibinfo {author} {\bibfnamefont {N.}~\bibnamefont {Itagaki}}, \ and\
  \bibinfo {author} {\bibfnamefont {J.}~\bibnamefont {Meng}},\ }\href {\doibase
  10.1103/PhysRevC.90.054307} {\bibfield  {journal} {\bibinfo  {journal} {Phys.
  Rev. C}\ }\textbf {\bibinfo {volume} {90}},\ \bibinfo {pages} {054307}
  (\bibinfo {year} {2014})}\BibitemShut {NoStop}%
\bibitem [{\citenamefont {Chevallier}\ \emph {et~al.}(1967)\citenamefont
  {Chevallier}, \citenamefont {Scheibling}, \citenamefont {Goldring},
  \citenamefont {Plesser},\ and\ \citenamefont {Sachs}}]{Chevallier1967O16LCS}%
  \BibitemOpen
  \bibfield  {author} {\bibinfo {author} {\bibfnamefont {P.}~\bibnamefont
  {Chevallier}}, \bibinfo {author} {\bibfnamefont {F.}~\bibnamefont
  {Scheibling}}, \bibinfo {author} {\bibfnamefont {G.}~\bibnamefont
  {Goldring}}, \bibinfo {author} {\bibfnamefont {I.}~\bibnamefont {Plesser}}, \
  and\ \bibinfo {author} {\bibfnamefont {M.~W.}\ \bibnamefont {Sachs}},\ }\href
  {\doibase 10.1103/PhysRev.160.827} {\bibfield  {journal} {\bibinfo  {journal}
  {Phys. Rev.}\ }\textbf {\bibinfo {volume} {160}},\ \bibinfo {pages} {827}
  (\bibinfo {year} {1967})}\BibitemShut {NoStop}%
\bibitem [{\citenamefont {Ichikawa}\ \emph {et~al.}(2011)\citenamefont
  {Ichikawa}, \citenamefont {Maruhn}, \citenamefont {Itagaki},\ and\
  \citenamefont {Ohkubo}}]{Ichikawa2011O16LinearChain}%
  \BibitemOpen
  \bibfield  {author} {\bibinfo {author} {\bibfnamefont {T.}~\bibnamefont
  {Ichikawa}}, \bibinfo {author} {\bibfnamefont {J.~A.}\ \bibnamefont
  {Maruhn}}, \bibinfo {author} {\bibfnamefont {N.}~\bibnamefont {Itagaki}}, \
  and\ \bibinfo {author} {\bibfnamefont {S.}~\bibnamefont {Ohkubo}},\ }\href
  {\doibase 10.1103/PhysRevLett.107.112501} {\bibfield  {journal} {\bibinfo
  {journal} {Phys. Rev. Lett.}\ }\textbf {\bibinfo {volume} {107}},\ \bibinfo
  {pages} {112501} (\bibinfo {year} {2011})}\BibitemShut {NoStop}%
\bibitem [{\citenamefont {Wuosmaa}\ \emph {et~al.}(1992)\citenamefont
  {Wuosmaa}, \citenamefont {Betts}, \citenamefont {Back}, \citenamefont
  {Freer}, \citenamefont {Glagola}, \citenamefont {Happ}, \citenamefont
  {Henderson}, \citenamefont {Wilt},\ and\ \citenamefont
  {Bearden}}]{Wuosmaa1992Mg24LCS}%
  \BibitemOpen
  \bibfield  {author} {\bibinfo {author} {\bibfnamefont {A.~H.}\ \bibnamefont
  {Wuosmaa}}, \bibinfo {author} {\bibfnamefont {R.~R.}\ \bibnamefont {Betts}},
  \bibinfo {author} {\bibfnamefont {B.~B.}\ \bibnamefont {Back}}, \bibinfo
  {author} {\bibfnamefont {M.}~\bibnamefont {Freer}}, \bibinfo {author}
  {\bibfnamefont {B.~G.}\ \bibnamefont {Glagola}}, \bibinfo {author}
  {\bibfnamefont {T.}~\bibnamefont {Happ}}, \bibinfo {author} {\bibfnamefont
  {D.~J.}\ \bibnamefont {Henderson}}, \bibinfo {author} {\bibfnamefont
  {P.}~\bibnamefont {Wilt}}, \ and\ \bibinfo {author} {\bibfnamefont {I.~G.}\
  \bibnamefont {Bearden}},\ }\href {\doibase 10.1103/PhysRevLett.68.1295}
  {\bibfield  {journal} {\bibinfo  {journal} {Phys. Rev. Lett.}\ }\textbf
  {\bibinfo {volume} {68}},\ \bibinfo {pages} {1295} (\bibinfo {year}
  {1992})}\BibitemShut {NoStop}%
\bibitem [{\citenamefont {Rae}\ \emph {et~al.}(1992)\citenamefont {Rae},
  \citenamefont {Merchant},\ and\ \citenamefont {Buck}}]{Rae1992Mg24}%
  \BibitemOpen
  \bibfield  {author} {\bibinfo {author} {\bibfnamefont {W.~D.~M.}\
  \bibnamefont {Rae}}, \bibinfo {author} {\bibfnamefont {A.~C.}\ \bibnamefont
  {Merchant}}, \ and\ \bibinfo {author} {\bibfnamefont {B.}~\bibnamefont
  {Buck}},\ }\href {\doibase 10.1103/PhysRevLett.69.3709} {\bibfield  {journal}
  {\bibinfo  {journal} {Phys. Rev. Lett.}\ }\textbf {\bibinfo {volume} {69}},\
  \bibinfo {pages} {3709} (\bibinfo {year} {1992})}\BibitemShut {NoStop}%
\bibitem [{\citenamefont {Iwata}\ \emph {et~al.}(2015)\citenamefont {Iwata},
  \citenamefont {Ichikawa}, \citenamefont {Itagaki}, \citenamefont {Maruhn},\
  and\ \citenamefont {Otsuka}}]{Iwata2015PRC_LCS_Mg}%
  \BibitemOpen
  \bibfield  {author} {\bibinfo {author} {\bibfnamefont {Y.}~\bibnamefont
  {Iwata}}, \bibinfo {author} {\bibfnamefont {T.}~\bibnamefont {Ichikawa}},
  \bibinfo {author} {\bibfnamefont {N.}~\bibnamefont {Itagaki}}, \bibinfo
  {author} {\bibfnamefont {J.~A.}\ \bibnamefont {Maruhn}}, \ and\ \bibinfo
  {author} {\bibfnamefont {T.}~\bibnamefont {Otsuka}},\ }\href {\doibase
  10.1103/PhysRevC.92.011303} {\bibfield  {journal} {\bibinfo  {journal} {Phys.
  Rev. C}\ }\textbf {\bibinfo {volume} {92}},\ \bibinfo {pages} {011303(R)}
  (\bibinfo {year} {2015})}\BibitemShut {NoStop}%
\bibitem [{\citenamefont {Inakura}\ and\ \citenamefont
  {Mizutori}(2018)}]{Inakura2018Rod_shaped}%
  \BibitemOpen
  \bibfield  {author} {\bibinfo {author} {\bibfnamefont {T.}~\bibnamefont
  {Inakura}}\ and\ \bibinfo {author} {\bibfnamefont {S.}~\bibnamefont
  {Mizutori}},\ }\href {\doibase 10.1103/PhysRevC.98.044312} {\bibfield
  {journal} {\bibinfo  {journal} {Phys. Rev. C}\ }\textbf {\bibinfo {volume}
  {98}},\ \bibinfo {pages} {044312} (\bibinfo {year} {2018})}\BibitemShut
  {NoStop}%
\bibitem [{\citenamefont {Itagaki}\ \emph {et~al.}(2001)\citenamefont
  {Itagaki}, \citenamefont {Okabe}, \citenamefont {Ikeda},\ and\ \citenamefont
  {Tanihata}}]{Itagaki2001MolecularOrbit}%
  \BibitemOpen
  \bibfield  {author} {\bibinfo {author} {\bibfnamefont {N.}~\bibnamefont
  {Itagaki}}, \bibinfo {author} {\bibfnamefont {S.}~\bibnamefont {Okabe}},
  \bibinfo {author} {\bibfnamefont {K.}~\bibnamefont {Ikeda}}, \ and\ \bibinfo
  {author} {\bibfnamefont {I.}~\bibnamefont {Tanihata}},\ }\href {\doibase
  10.1103/PhysRevC.64.014301} {\bibfield  {journal} {\bibinfo  {journal} {Phys.
  Rev. C}\ }\textbf {\bibinfo {volume} {64}},\ \bibinfo {pages} {014301}
  (\bibinfo {year} {2001})}\BibitemShut {NoStop}%
\bibitem [{\citenamefont {Freer}\ \emph {et~al.}(2006)\citenamefont {Freer},
  \citenamefont {Casarejos}, \citenamefont {Achouri}, \citenamefont {Angulo},
  \citenamefont {Ashwood}, \citenamefont {Curtis}, \citenamefont {Demaret},
  \citenamefont {Harlin}, \citenamefont {Laurent}, \citenamefont {Milin},
  \citenamefont {Orr}, \citenamefont {Price}, \citenamefont {Raabe},
  \citenamefont {Soi\ifmmode~\acute{c}\else \'{c}\fi{}},\ and\ \citenamefont
  {Ziman}}]{Freer2006Phys.Rev.Lett.42501}%
  \BibitemOpen
  \bibfield  {author} {\bibinfo {author} {\bibfnamefont {M.}~\bibnamefont
  {Freer}}, \bibinfo {author} {\bibfnamefont {E.}~\bibnamefont {Casarejos}},
  \bibinfo {author} {\bibfnamefont {L.}~\bibnamefont {Achouri}}, \bibinfo
  {author} {\bibfnamefont {C.}~\bibnamefont {Angulo}}, \bibinfo {author}
  {\bibfnamefont {N.~I.}\ \bibnamefont {Ashwood}}, \bibinfo {author}
  {\bibfnamefont {N.}~\bibnamefont {Curtis}}, \bibinfo {author} {\bibfnamefont
  {P.}~\bibnamefont {Demaret}}, \bibinfo {author} {\bibfnamefont
  {C.}~\bibnamefont {Harlin}}, \bibinfo {author} {\bibfnamefont
  {B.}~\bibnamefont {Laurent}}, \bibinfo {author} {\bibfnamefont
  {M.}~\bibnamefont {Milin}}, \bibinfo {author} {\bibfnamefont {N.~A.}\
  \bibnamefont {Orr}}, \bibinfo {author} {\bibfnamefont {D.}~\bibnamefont
  {Price}}, \bibinfo {author} {\bibfnamefont {R.}~\bibnamefont {Raabe}},
  \bibinfo {author} {\bibfnamefont {N.}~\bibnamefont
  {Soi\ifmmode~\acute{c}\else \'{c}\fi{}}}, \ and\ \bibinfo {author}
  {\bibfnamefont {V.~A.}\ \bibnamefont {Ziman}},\ }\href {\doibase
  10.1103/PhysRevLett.96.042501} {\bibfield  {journal} {\bibinfo  {journal}
  {Phys. Rev. Lett.}\ }\textbf {\bibinfo {volume} {96}},\ \bibinfo {pages}
  {042501} (\bibinfo {year} {2006})}\BibitemShut {NoStop}%
\bibitem [{\citenamefont {Navin}\ \emph {et~al.}(2000)\citenamefont {Navin},
  \citenamefont {Anthony}, \citenamefont {Aumann}, \citenamefont {Baumann},
  \citenamefont {Bazin}, \citenamefont {Blumenfeld}, \citenamefont {Brown},
  \citenamefont {Glasmacher}, \citenamefont {Hansen}, \citenamefont {Ibbotson},
  \citenamefont {Lofy}, \citenamefont {Maddalena}, \citenamefont {Miller},
  \citenamefont {Nakamura}, \citenamefont {Pritychenko}, \citenamefont
  {Sherrill}, \citenamefont {Spears}, \citenamefont {Steiner}, \citenamefont
  {Tostevin}, \citenamefont {Yurkon},\ and\ \citenamefont
  {Wagner}}]{Navin2000Phys.Rev.Lett.266}%
  \BibitemOpen
  \bibfield  {author} {\bibinfo {author} {\bibfnamefont {A.}~\bibnamefont
  {Navin}}, \bibinfo {author} {\bibfnamefont {D.~W.}\ \bibnamefont {Anthony}},
  \bibinfo {author} {\bibfnamefont {T.}~\bibnamefont {Aumann}}, \bibinfo
  {author} {\bibfnamefont {T.}~\bibnamefont {Baumann}}, \bibinfo {author}
  {\bibfnamefont {D.}~\bibnamefont {Bazin}}, \bibinfo {author} {\bibfnamefont
  {Y.}~\bibnamefont {Blumenfeld}}, \bibinfo {author} {\bibfnamefont {B.~A.}\
  \bibnamefont {Brown}}, \bibinfo {author} {\bibfnamefont {T.}~\bibnamefont
  {Glasmacher}}, \bibinfo {author} {\bibfnamefont {P.~G.}\ \bibnamefont
  {Hansen}}, \bibinfo {author} {\bibfnamefont {R.~W.}\ \bibnamefont
  {Ibbotson}}, \bibinfo {author} {\bibfnamefont {P.~A.}\ \bibnamefont {Lofy}},
  \bibinfo {author} {\bibfnamefont {V.}~\bibnamefont {Maddalena}}, \bibinfo
  {author} {\bibfnamefont {K.}~\bibnamefont {Miller}}, \bibinfo {author}
  {\bibfnamefont {T.}~\bibnamefont {Nakamura}}, \bibinfo {author}
  {\bibfnamefont {B.~V.}\ \bibnamefont {Pritychenko}}, \bibinfo {author}
  {\bibfnamefont {B.~M.}\ \bibnamefont {Sherrill}}, \bibinfo {author}
  {\bibfnamefont {E.}~\bibnamefont {Spears}}, \bibinfo {author} {\bibfnamefont
  {M.}~\bibnamefont {Steiner}}, \bibinfo {author} {\bibfnamefont {J.~A.}\
  \bibnamefont {Tostevin}}, \bibinfo {author} {\bibfnamefont {J.}~\bibnamefont
  {Yurkon}}, \ and\ \bibinfo {author} {\bibfnamefont {A.}~\bibnamefont
  {Wagner}},\ }\href {\doibase 10.1103/PhysRevLett.85.266} {\bibfield
  {journal} {\bibinfo  {journal} {Phys. Rev. Lett.}\ }\textbf {\bibinfo
  {volume} {85}},\ \bibinfo {pages} {266} (\bibinfo {year} {2000})}\BibitemShut
  {NoStop}%
\bibitem [{\citenamefont {Itagaki}\ and\ \citenamefont
  {Okabe}(2000)}]{Itagaki2000Phys.Rev.C44306}%
  \BibitemOpen
  \bibfield  {author} {\bibinfo {author} {\bibfnamefont {N.}~\bibnamefont
  {Itagaki}}\ and\ \bibinfo {author} {\bibfnamefont {S.}~\bibnamefont
  {Okabe}},\ }\href {\doibase 10.1103/PhysRevC.61.044306} {\bibfield  {journal}
  {\bibinfo  {journal} {Phys. Rev. C}\ }\textbf {\bibinfo {volume} {61}},\
  \bibinfo {pages} {044306} (\bibinfo {year} {2000})}\BibitemShut {NoStop}%
\bibitem [{\citenamefont {von Oertzen}\ \emph {et~al.}(2004)\citenamefont {von
  Oertzen}, \citenamefont {Bohlen}, \citenamefont {Milin}, \citenamefont
  {Kokalova}, \citenamefont {Thummerer}, \citenamefont {Tumino}, \citenamefont
  {Kalpakchieva}, \citenamefont {Massey}, \citenamefont {Eisermann},
  \citenamefont {Graw}, \citenamefont {Faestermann}, \citenamefont
  {Hertenberger},\ and\ \citenamefont {Wirth}}]{Oertzenn2004C14_LCS}%
  \BibitemOpen
  \bibfield  {author} {\bibinfo {author} {\bibfnamefont {W.}~\bibnamefont {von
  Oertzen}}, \bibinfo {author} {\bibfnamefont {H.~G.}\ \bibnamefont {Bohlen}},
  \bibinfo {author} {\bibfnamefont {M.}~\bibnamefont {Milin}}, \bibinfo
  {author} {\bibfnamefont {T.}~\bibnamefont {Kokalova}}, \bibinfo {author}
  {\bibfnamefont {S.}~\bibnamefont {Thummerer}}, \bibinfo {author}
  {\bibfnamefont {A.}~\bibnamefont {Tumino}}, \bibinfo {author} {\bibfnamefont
  {R.}~\bibnamefont {Kalpakchieva}}, \bibinfo {author} {\bibfnamefont {T.~N.}\
  \bibnamefont {Massey}}, \bibinfo {author} {\bibfnamefont {Y.}~\bibnamefont
  {Eisermann}}, \bibinfo {author} {\bibfnamefont {G.}~\bibnamefont {Graw}},
  \bibinfo {author} {\bibfnamefont {T.}~\bibnamefont {Faestermann}}, \bibinfo
  {author} {\bibfnamefont {R.}~\bibnamefont {Hertenberger}}, \ and\ \bibinfo
  {author} {\bibfnamefont {H.-F.}\ \bibnamefont {Wirth}},\ }\href {\doibase
  10.1140/epja/i2003-10188-9} {\bibfield  {journal} {\bibinfo  {journal} {Eur.
  Phys. J. A}\ }\textbf {\bibinfo {volume} {21}},\ \bibinfo {pages} {193}
  (\bibinfo {year} {2004})}\BibitemShut {NoStop}%
\bibitem [{\citenamefont {Ito}\ \emph {et~al.}(2008)\citenamefont {Ito},
  \citenamefont {Itagaki}, \citenamefont {Sakurai},\ and\ \citenamefont
  {Ikeda}}]{Ito2008Phys.Rev.Lett.182502}%
  \BibitemOpen
  \bibfield  {author} {\bibinfo {author} {\bibfnamefont {M.}~\bibnamefont
  {Ito}}, \bibinfo {author} {\bibfnamefont {N.}~\bibnamefont {Itagaki}},
  \bibinfo {author} {\bibfnamefont {H.}~\bibnamefont {Sakurai}}, \ and\
  \bibinfo {author} {\bibfnamefont {K.}~\bibnamefont {Ikeda}},\ }\href
  {\doibase 10.1103/PhysRevLett.100.182502} {\bibfield  {journal} {\bibinfo
  {journal} {Phys. Rev. Lett.}\ }\textbf {\bibinfo {volume} {100}},\ \bibinfo
  {pages} {182502} (\bibinfo {year} {2008})}\BibitemShut {NoStop}%
\bibitem [{\citenamefont {Maruhn}\ \emph {et~al.}(2010)\citenamefont {Maruhn},
  \citenamefont {Loebl}, \citenamefont {Itagaki},\ and\ \citenamefont
  {Kimura}}]{Maruhn2010NPA_LCS_Carbon}%
  \BibitemOpen
  \bibfield  {author} {\bibinfo {author} {\bibfnamefont {J.~A.}\ \bibnamefont
  {Maruhn}}, \bibinfo {author} {\bibfnamefont {N.}~\bibnamefont {Loebl}},
  \bibinfo {author} {\bibfnamefont {N.}~\bibnamefont {Itagaki}}, \ and\
  \bibinfo {author} {\bibfnamefont {M.}~\bibnamefont {Kimura}},\ }\href
  {\doibase https://doi.org/10.1016/j.nuclphysa.2009.12.005} {\bibfield
  {journal} {\bibinfo  {journal} {Nucl. Phys. A}\ }\textbf {\bibinfo {volume}
  {833}},\ \bibinfo {pages} {1} (\bibinfo {year} {2010})}\BibitemShut {NoStop}%
\bibitem [{\citenamefont {Baba}\ \emph {et~al.}(2014)\citenamefont {Baba},
  \citenamefont {Chiba},\ and\ \citenamefont
  {Kimura}}]{Baba2014Phys.Rev.C64319}%
  \BibitemOpen
  \bibfield  {author} {\bibinfo {author} {\bibfnamefont {T.}~\bibnamefont
  {Baba}}, \bibinfo {author} {\bibfnamefont {Y.}~\bibnamefont {Chiba}}, \ and\
  \bibinfo {author} {\bibfnamefont {M.}~\bibnamefont {Kimura}},\ }\href
  {\doibase 10.1103/PhysRevC.90.064319} {\bibfield  {journal} {\bibinfo
  {journal} {Phys. Rev. C}\ }\textbf {\bibinfo {volume} {90}},\ \bibinfo
  {pages} {064319} (\bibinfo {year} {2014})}\BibitemShut {NoStop}%
\bibitem [{\citenamefont {Freer}\ \emph {et~al.}(2011)\citenamefont {Freer},
  \citenamefont {Ashwood}, \citenamefont {Curtis}, \citenamefont {Di~Pietro},
  \citenamefont {Figuera}, \citenamefont {Fisichella}, \citenamefont {Grassi},
  \citenamefont {Jelavi\ifmmode \acute{c}\else~\'{c}\fi{} Malenica},
  \citenamefont {Kokalova}, \citenamefont {Koncul}, \citenamefont
  {Mijatovi\ifmmode~\acute{c}\else \'{c}\fi{}}, \citenamefont {Milin},
  \citenamefont {Prepolec}, \citenamefont {Scuderi}, \citenamefont {Skukan},
  \citenamefont {Soi\ifmmode~\acute{c}\else \'{c}\fi{}}, \citenamefont
  {Szilner}, \citenamefont {Toki\ifmmode~\acute{c}\else \'{c}\fi{}},
  \citenamefont {Torresi},\ and\ \citenamefont {Wheldon}}]{Freer2011PRC13C}%
  \BibitemOpen
  \bibfield  {author} {\bibinfo {author} {\bibfnamefont {M.}~\bibnamefont
  {Freer}}, \bibinfo {author} {\bibfnamefont {N.~I.}\ \bibnamefont {Ashwood}},
  \bibinfo {author} {\bibfnamefont {N.}~\bibnamefont {Curtis}}, \bibinfo
  {author} {\bibfnamefont {A.}~\bibnamefont {Di~Pietro}}, \bibinfo {author}
  {\bibfnamefont {P.}~\bibnamefont {Figuera}}, \bibinfo {author} {\bibfnamefont
  {M.}~\bibnamefont {Fisichella}}, \bibinfo {author} {\bibfnamefont
  {L.}~\bibnamefont {Grassi}}, \bibinfo {author} {\bibfnamefont
  {D.}~\bibnamefont {Jelavi\ifmmode \acute{c}\else~\'{c}\fi{} Malenica}},
  \bibinfo {author} {\bibfnamefont {T.}~\bibnamefont {Kokalova}}, \bibinfo
  {author} {\bibfnamefont {M.}~\bibnamefont {Koncul}}, \bibinfo {author}
  {\bibfnamefont {T.}~\bibnamefont {Mijatovi\ifmmode~\acute{c}\else
  \'{c}\fi{}}}, \bibinfo {author} {\bibfnamefont {M.}~\bibnamefont {Milin}},
  \bibinfo {author} {\bibfnamefont {L.}~\bibnamefont {Prepolec}}, \bibinfo
  {author} {\bibfnamefont {V.}~\bibnamefont {Scuderi}}, \bibinfo {author}
  {\bibfnamefont {N.}~\bibnamefont {Skukan}}, \bibinfo {author} {\bibfnamefont
  {N.}~\bibnamefont {Soi\ifmmode~\acute{c}\else \'{c}\fi{}}}, \bibinfo {author}
  {\bibfnamefont {S.}~\bibnamefont {Szilner}}, \bibinfo {author} {\bibfnamefont
  {V.}~\bibnamefont {Toki\ifmmode~\acute{c}\else \'{c}\fi{}}}, \bibinfo
  {author} {\bibfnamefont {D.}~\bibnamefont {Torresi}}, \ and\ \bibinfo
  {author} {\bibfnamefont {C.}~\bibnamefont {Wheldon}},\ }\href {\doibase
  10.1103/PhysRevC.84.034317} {\bibfield  {journal} {\bibinfo  {journal} {Phys.
  Rev. C}\ }\textbf {\bibinfo {volume} {84}},\ \bibinfo {pages} {034317}
  (\bibinfo {year} {2011})}\BibitemShut {NoStop}%
\bibitem [{\citenamefont {Freer}\ \emph {et~al.}(2014)\citenamefont {Freer},
  \citenamefont {Malcolm}, \citenamefont {Achouri}, \citenamefont {Ashwood},
  \citenamefont {Bardayan}, \citenamefont {Brown}, \citenamefont {Catford},
  \citenamefont {Chipps}, \citenamefont {Cizewski}, \citenamefont {Curtis},
  \citenamefont {Jones}, \citenamefont {Munoz-Britton}, \citenamefont {Pain},
  \citenamefont {Soi\ifmmode~\acute{c}\else \'{c}\fi{}}, \citenamefont
  {Wheldon}, \citenamefont {Wilson},\ and\ \citenamefont
  {Ziman}}]{Freer2014C14_LCS_C14}%
  \BibitemOpen
  \bibfield  {author} {\bibinfo {author} {\bibfnamefont {M.}~\bibnamefont
  {Freer}}, \bibinfo {author} {\bibfnamefont {J.~D.}\ \bibnamefont {Malcolm}},
  \bibinfo {author} {\bibfnamefont {N.~L.}\ \bibnamefont {Achouri}}, \bibinfo
  {author} {\bibfnamefont {N.~I.}\ \bibnamefont {Ashwood}}, \bibinfo {author}
  {\bibfnamefont {D.~W.}\ \bibnamefont {Bardayan}}, \bibinfo {author}
  {\bibfnamefont {S.~M.}\ \bibnamefont {Brown}}, \bibinfo {author}
  {\bibfnamefont {W.~N.}\ \bibnamefont {Catford}}, \bibinfo {author}
  {\bibfnamefont {K.~A.}\ \bibnamefont {Chipps}}, \bibinfo {author}
  {\bibfnamefont {J.}~\bibnamefont {Cizewski}}, \bibinfo {author}
  {\bibfnamefont {N.}~\bibnamefont {Curtis}}, \bibinfo {author} {\bibfnamefont
  {K.~L.}\ \bibnamefont {Jones}}, \bibinfo {author} {\bibfnamefont
  {T.}~\bibnamefont {Munoz-Britton}}, \bibinfo {author} {\bibfnamefont {S.~D.}\
  \bibnamefont {Pain}}, \bibinfo {author} {\bibfnamefont {N.}~\bibnamefont
  {Soi\ifmmode~\acute{c}\else \'{c}\fi{}}}, \bibinfo {author} {\bibfnamefont
  {C.}~\bibnamefont {Wheldon}}, \bibinfo {author} {\bibfnamefont {G.~L.}\
  \bibnamefont {Wilson}}, \ and\ \bibinfo {author} {\bibfnamefont {V.~A.}\
  \bibnamefont {Ziman}},\ }\href {\doibase 10.1103/PhysRevC.90.054324}
  {\bibfield  {journal} {\bibinfo  {journal} {Phys. Rev. C}\ }\textbf {\bibinfo
  {volume} {90}},\ \bibinfo {pages} {054324} (\bibinfo {year}
  {2014})}\BibitemShut {NoStop}%
\bibitem [{\citenamefont {Fritsch}\ \emph {et~al.}(2016)\citenamefont
  {Fritsch}, \citenamefont {Beceiro-Novo}, \citenamefont {Suzuki},
  \citenamefont {Mittig}, \citenamefont {Kolata}, \citenamefont {Ahn},
  \citenamefont {Bazin}, \citenamefont {Becchetti}, \citenamefont {Bucher},
  \citenamefont {Chajecki}, \citenamefont {Fang}, \citenamefont {Febbraro},
  \citenamefont {Howard}, \citenamefont {Kanada-En'yo}, \citenamefont {Lynch},
  \citenamefont {Mitchell}, \citenamefont {Ojaruega}, \citenamefont {Rogers},
  \citenamefont {Shore}, \citenamefont {Suhara}, \citenamefont {Tang},
  \citenamefont {Torres-Isea},\ and\ \citenamefont {Wang}}]{Fritsch2016C14LCS}%
  \BibitemOpen
  \bibfield  {author} {\bibinfo {author} {\bibfnamefont {A.}~\bibnamefont
  {Fritsch}}, \bibinfo {author} {\bibfnamefont {S.}~\bibnamefont
  {Beceiro-Novo}}, \bibinfo {author} {\bibfnamefont {D.}~\bibnamefont
  {Suzuki}}, \bibinfo {author} {\bibfnamefont {W.}~\bibnamefont {Mittig}},
  \bibinfo {author} {\bibfnamefont {J.~J.}\ \bibnamefont {Kolata}}, \bibinfo
  {author} {\bibfnamefont {T.}~\bibnamefont {Ahn}}, \bibinfo {author}
  {\bibfnamefont {D.}~\bibnamefont {Bazin}}, \bibinfo {author} {\bibfnamefont
  {F.~D.}\ \bibnamefont {Becchetti}}, \bibinfo {author} {\bibfnamefont
  {B.}~\bibnamefont {Bucher}}, \bibinfo {author} {\bibfnamefont
  {Z.}~\bibnamefont {Chajecki}}, \bibinfo {author} {\bibfnamefont
  {X.}~\bibnamefont {Fang}}, \bibinfo {author} {\bibfnamefont {M.}~\bibnamefont
  {Febbraro}}, \bibinfo {author} {\bibfnamefont {A.~M.}\ \bibnamefont
  {Howard}}, \bibinfo {author} {\bibfnamefont {Y.}~\bibnamefont
  {Kanada-En'yo}}, \bibinfo {author} {\bibfnamefont {W.~G.}\ \bibnamefont
  {Lynch}}, \bibinfo {author} {\bibfnamefont {A.~J.}\ \bibnamefont {Mitchell}},
  \bibinfo {author} {\bibfnamefont {M.}~\bibnamefont {Ojaruega}}, \bibinfo
  {author} {\bibfnamefont {A.~M.}\ \bibnamefont {Rogers}}, \bibinfo {author}
  {\bibfnamefont {A.}~\bibnamefont {Shore}}, \bibinfo {author} {\bibfnamefont
  {T.}~\bibnamefont {Suhara}}, \bibinfo {author} {\bibfnamefont {X.~D.}\
  \bibnamefont {Tang}}, \bibinfo {author} {\bibfnamefont {R.}~\bibnamefont
  {Torres-Isea}}, \ and\ \bibinfo {author} {\bibfnamefont {H.}~\bibnamefont
  {Wang}},\ }\href {\doibase 10.1103/PhysRevC.93.014321} {\bibfield  {journal}
  {\bibinfo  {journal} {Phys. Rev. C}\ }\textbf {\bibinfo {volume} {93}},\
  \bibinfo {pages} {014321} (\bibinfo {year} {2016})}\BibitemShut {NoStop}%
\bibitem [{\citenamefont {Yamaguchi}\ \emph {et~al.}(2017)\citenamefont
  {Yamaguchi}, \citenamefont {Kahl}, \citenamefont {Hayakawa}, \citenamefont
  {Sakaguchi}, \citenamefont {Abe}, \citenamefont {Nakao}, \citenamefont
  {Suhara}, \citenamefont {Iwasa}, \citenamefont {Kim}, \citenamefont {Kim},
  \citenamefont {Cha}, \citenamefont {Kwag}, \citenamefont {Lee}, \citenamefont
  {Lee}, \citenamefont {Chae}, \citenamefont {Wakabayashi}, \citenamefont
  {Imai}, \citenamefont {Kitamura}, \citenamefont {Lee}, \citenamefont {Moon},
  \citenamefont {Lee}, \citenamefont {Akers}, \citenamefont {Jung},
  \citenamefont {Duy}, \citenamefont {Khiem},\ and\ \citenamefont
  {Lee}}]{Yamaguchi2017C14LCS}%
  \BibitemOpen
  \bibfield  {author} {\bibinfo {author} {\bibfnamefont {H.}~\bibnamefont
  {Yamaguchi}}, \bibinfo {author} {\bibfnamefont {D.}~\bibnamefont {Kahl}},
  \bibinfo {author} {\bibfnamefont {S.}~\bibnamefont {Hayakawa}}, \bibinfo
  {author} {\bibfnamefont {Y.}~\bibnamefont {Sakaguchi}}, \bibinfo {author}
  {\bibfnamefont {K.}~\bibnamefont {Abe}}, \bibinfo {author} {\bibfnamefont
  {T.}~\bibnamefont {Nakao}}, \bibinfo {author} {\bibfnamefont
  {T.}~\bibnamefont {Suhara}}, \bibinfo {author} {\bibfnamefont
  {N.}~\bibnamefont {Iwasa}}, \bibinfo {author} {\bibfnamefont
  {A.}~\bibnamefont {Kim}}, \bibinfo {author} {\bibfnamefont {D.~H.}\
  \bibnamefont {Kim}}, \bibinfo {author} {\bibfnamefont {S.~M.}\ \bibnamefont
  {Cha}}, \bibinfo {author} {\bibfnamefont {M.~S.}\ \bibnamefont {Kwag}},
  \bibinfo {author} {\bibfnamefont {J.~H.}\ \bibnamefont {Lee}}, \bibinfo
  {author} {\bibfnamefont {E.~J.}\ \bibnamefont {Lee}}, \bibinfo {author}
  {\bibfnamefont {K.~Y.}\ \bibnamefont {Chae}}, \bibinfo {author}
  {\bibfnamefont {Y.}~\bibnamefont {Wakabayashi}}, \bibinfo {author}
  {\bibfnamefont {N.}~\bibnamefont {Imai}}, \bibinfo {author} {\bibfnamefont
  {N.}~\bibnamefont {Kitamura}}, \bibinfo {author} {\bibfnamefont
  {P.}~\bibnamefont {Lee}}, \bibinfo {author} {\bibfnamefont {J.~Y.}\
  \bibnamefont {Moon}}, \bibinfo {author} {\bibfnamefont {K.~B.}\ \bibnamefont
  {Lee}}, \bibinfo {author} {\bibfnamefont {C.}~\bibnamefont {Akers}}, \bibinfo
  {author} {\bibfnamefont {H.~S.}\ \bibnamefont {Jung}}, \bibinfo {author}
  {\bibfnamefont {N.~N.}\ \bibnamefont {Duy}}, \bibinfo {author} {\bibfnamefont
  {L.~H.}\ \bibnamefont {Khiem}}, \ and\ \bibinfo {author} {\bibfnamefont
  {C.~S.}\ \bibnamefont {Lee}},\ }\href {\doibase
  https://doi.org/10.1016/j.physletb.2016.12.050} {\bibfield  {journal}
  {\bibinfo  {journal} {Phys. Lett. B}\ }\textbf {\bibinfo {volume} {766}},\
  \bibinfo {pages} {11} (\bibinfo {year} {2017})}\BibitemShut {NoStop}%
\bibitem [{\citenamefont {Suhara}\ and\ \citenamefont
  {Kanada-En'yo}(2010)}]{Suhara2010C14ClusterExcited}%
  \BibitemOpen
  \bibfield  {author} {\bibinfo {author} {\bibfnamefont {T.}~\bibnamefont
  {Suhara}}\ and\ \bibinfo {author} {\bibfnamefont {Y.}~\bibnamefont
  {Kanada-En'yo}},\ }\href {\doibase 10.1103/PhysRevC.82.044301} {\bibfield
  {journal} {\bibinfo  {journal} {Phys. Rev. C}\ }\textbf {\bibinfo {volume}
  {82}},\ \bibinfo {pages} {044301} (\bibinfo {year} {2010})}\BibitemShut
  {NoStop}%
\bibitem [{\citenamefont {Negele}(1982)}]{Negele1982TDDFT}%
  \BibitemOpen
  \bibfield  {author} {\bibinfo {author} {\bibfnamefont {J.~W.}\ \bibnamefont
  {Negele}},\ }\href {\doibase 10.1103/RevModPhys.54.913} {\bibfield  {journal}
  {\bibinfo  {journal} {Rev. Mod. Phys.}\ }\textbf {\bibinfo {volume} {54}},\
  \bibinfo {pages} {913} (\bibinfo {year} {1982})}\BibitemShut {NoStop}%
\bibitem [{\citenamefont {Simenel}(2012)}]{Simene2012PEPJA}%
  \BibitemOpen
  \bibfield  {author} {\bibinfo {author} {\bibfnamefont {C.}~\bibnamefont
  {Simenel}},\ }\href {\doibase 10.1140/epja/i2012-12152-0} {\bibfield
  {journal} {\bibinfo  {journal} {Eur. Phys. J. A}\ }\textbf {\bibinfo {volume}
  {48}},\ \bibinfo {pages} {152} (\bibinfo {year} {2012})}\BibitemShut
  {NoStop}%
\bibitem [{\citenamefont {Nakatsukasa}\ \emph {et~al.}(2016)\citenamefont
  {Nakatsukasa}, \citenamefont {Matsuyanagi}, \citenamefont {Matsuo},\ and\
  \citenamefont {Yabana}}]{NakatsukasaRMP2016}%
  \BibitemOpen
  \bibfield  {author} {\bibinfo {author} {\bibfnamefont {T.}~\bibnamefont
  {Nakatsukasa}}, \bibinfo {author} {\bibfnamefont {K.}~\bibnamefont
  {Matsuyanagi}}, \bibinfo {author} {\bibfnamefont {M.}~\bibnamefont {Matsuo}},
  \ and\ \bibinfo {author} {\bibfnamefont {K.}~\bibnamefont {Yabana}},\ }\href
  {\doibase 10.1103/RevModPhys.88.045004} {\bibfield  {journal} {\bibinfo
  {journal} {Rev. Mod. Phys.}\ }\textbf {\bibinfo {volume} {88}},\ \bibinfo
  {pages} {045004} (\bibinfo {year} {2016})}\BibitemShut {NoStop}%
\bibitem [{\citenamefont {Simenel}\ and\ \citenamefont
  {Umar}(2018)}]{SIMENEL2018TDHF_PPNP}%
  \BibitemOpen
  \bibfield  {author} {\bibinfo {author} {\bibfnamefont {C.}~\bibnamefont
  {Simenel}}\ and\ \bibinfo {author} {\bibfnamefont {A.~S.}\ \bibnamefont
  {Umar}},\ }\href {\doibase https://doi.org/10.1016/j.ppnp.2018.07.002}
  {\bibfield  {journal} {\bibinfo  {journal} {Prog. Part. Nucl. Phys.}\
  }\textbf {\bibinfo {volume} {103}},\ \bibinfo {pages} {19} (\bibinfo {year}
  {2018})}\BibitemShut {NoStop}%
\bibitem [{\citenamefont {Stevenson}\ and\ \citenamefont
  {Barton}(2019)}]{STEVENSON2019PPNP}%
  \BibitemOpen
  \bibfield  {author} {\bibinfo {author} {\bibfnamefont {P.~D.}\ \bibnamefont
  {Stevenson}}\ and\ \bibinfo {author} {\bibfnamefont {M.~C.}\ \bibnamefont
  {Barton}},\ }\href {\doibase https://doi.org/10.1016/j.ppnp.2018.09.002}
  {\bibfield  {journal} {\bibinfo  {journal} {Prog. Part. Nucl. Phys.}\
  }\textbf {\bibinfo {volume} {104}},\ \bibinfo {pages} {142} (\bibinfo {year}
  {2019})}\BibitemShut {NoStop}%
\bibitem [{\citenamefont {Umar}\ \emph {et~al.}(2010)\citenamefont {Umar},
  \citenamefont {Maruhn}, \citenamefont {Itagaki},\ and\ \citenamefont
  {Oberacker}}]{Umar2010TDHF_C12}%
  \BibitemOpen
  \bibfield  {author} {\bibinfo {author} {\bibfnamefont {A.~S.}\ \bibnamefont
  {Umar}}, \bibinfo {author} {\bibfnamefont {J.~A.}\ \bibnamefont {Maruhn}},
  \bibinfo {author} {\bibfnamefont {N.}~\bibnamefont {Itagaki}}, \ and\
  \bibinfo {author} {\bibfnamefont {V.~E.}\ \bibnamefont {Oberacker}},\ }\href
  {\doibase 10.1103/PhysRevLett.104.212503} {\bibfield  {journal} {\bibinfo
  {journal} {Phys. Rev. Lett.}\ }\textbf {\bibinfo {volume} {104}},\ \bibinfo
  {pages} {212503} (\bibinfo {year} {2010})}\BibitemShut {NoStop}%
\bibitem [{\citenamefont {Ring}(1996)}]{RING1996PPNP}%
  \BibitemOpen
  \bibfield  {author} {\bibinfo {author} {\bibfnamefont {P.}~\bibnamefont
  {Ring}},\ }\href {\doibase http://dx.doi.org/10.1016/0146-6410(96)00054-3}
  {\bibfield  {journal} {\bibinfo  {journal} {Prog. Part. Nucl. Phys.}\
  }\textbf {\bibinfo {volume} {37}},\ \bibinfo {pages} {193} (\bibinfo {year}
  {1996})}\BibitemShut {NoStop}%
\bibitem [{\citenamefont {Vretenar}\ \emph {et~al.}(2005)\citenamefont
  {Vretenar}, \citenamefont {Afanasjev}, \citenamefont {Lalazissis},\ and\
  \citenamefont {Ring}}]{Vretenar2005PhysicsReport}%
  \BibitemOpen
  \bibfield  {author} {\bibinfo {author} {\bibfnamefont {D.}~\bibnamefont
  {Vretenar}}, \bibinfo {author} {\bibfnamefont {A.~V.}\ \bibnamefont
  {Afanasjev}}, \bibinfo {author} {\bibfnamefont {G.~A.}\ \bibnamefont
  {Lalazissis}}, \ and\ \bibinfo {author} {\bibfnamefont {P.}~\bibnamefont
  {Ring}},\ }\href {\doibase http://dx.doi.org/10.1016/j.physrep.2004.10.001}
  {\bibfield  {journal} {\bibinfo  {journal} {Phys. Rep.}\ }\textbf {\bibinfo
  {volume} {409}},\ \bibinfo {pages} {101} (\bibinfo {year}
  {2005})}\BibitemShut {NoStop}%
\bibitem [{\citenamefont {Meng}\ \emph {et~al.}(2006)\citenamefont {Meng},
  \citenamefont {Toki}, \citenamefont {Zhou}, \citenamefont {Zhang},
  \citenamefont {Long},\ and\ \citenamefont {Geng}}]{meng2006PPNP}%
  \BibitemOpen
  \bibfield  {author} {\bibinfo {author} {\bibfnamefont {J.}~\bibnamefont
  {Meng}}, \bibinfo {author} {\bibfnamefont {H.}~\bibnamefont {Toki}}, \bibinfo
  {author} {\bibfnamefont {S.~G.}\ \bibnamefont {Zhou}}, \bibinfo {author}
  {\bibfnamefont {S.~Q.}\ \bibnamefont {Zhang}}, \bibinfo {author}
  {\bibfnamefont {W.~H.}\ \bibnamefont {Long}}, \ and\ \bibinfo {author}
  {\bibfnamefont {L.~S.}\ \bibnamefont {Geng}},\ }\href {\doibase
  https://doi.org/10.1016/j.ppnp.2005.06.001} {\bibfield  {journal} {\bibinfo
  {journal} {Prog. Part. Nucl. Phys.}\ }\textbf {\bibinfo {volume} {57}},\
  \bibinfo {pages} {470} (\bibinfo {year} {2006})}\BibitemShut {NoStop}%
\bibitem [{\citenamefont {Meng}\ \emph {et~al.}(2013)\citenamefont {Meng},
  \citenamefont {Peng}, \citenamefont {Zhang},\ and\ \citenamefont
  {Zhao}}]{Meng2013FT_TAC}%
  \BibitemOpen
  \bibfield  {author} {\bibinfo {author} {\bibfnamefont {J.}~\bibnamefont
  {Meng}}, \bibinfo {author} {\bibfnamefont {J.}~\bibnamefont {Peng}}, \bibinfo
  {author} {\bibfnamefont {S.-Q.}\ \bibnamefont {Zhang}}, \ and\ \bibinfo
  {author} {\bibfnamefont {P.-W.}\ \bibnamefont {Zhao}},\ }\href {\doibase
  10.1007/s11467-013-0287-y} {\bibfield  {journal} {\bibinfo  {journal} {Front.
  Phys.}\ }\textbf {\bibinfo {volume} {8}},\ \bibinfo {pages} {55} (\bibinfo
  {year} {2013})}\BibitemShut {NoStop}%
\bibitem [{\citenamefont {Meng}(2016)}]{meng2016relativistic}%
  \BibitemOpen
  \bibinfo {editor} {\bibfnamefont {J.}~\bibnamefont {Meng}},\ ed.,\ \href@noop
  {} {\emph {\bibinfo {title} {Relativistic Density Functional for Nuclear
  Structure}}},\ \bibinfo {series} {International Review of Nuclear Physics},
  Vol.~\bibinfo {volume} {10}\ (\bibinfo  {publisher} {World Scientific,
  Singapore},\ \bibinfo {year} {2016})\BibitemShut {NoStop}%
\bibitem [{\citenamefont {Cusson}\ \emph {et~al.}(1985)\citenamefont {Cusson},
  \citenamefont {Reinhard}, \citenamefont {Molitoris}, \citenamefont
  {St\"ocker}, \citenamefont {Strayer},\ and\ \citenamefont
  {Greiner}}]{Cusson1985TDCDFT}%
  \BibitemOpen
  \bibfield  {author} {\bibinfo {author} {\bibfnamefont {R.~Y.}\ \bibnamefont
  {Cusson}}, \bibinfo {author} {\bibfnamefont {P.~G.}\ \bibnamefont
  {Reinhard}}, \bibinfo {author} {\bibfnamefont {J.~J.}\ \bibnamefont
  {Molitoris}}, \bibinfo {author} {\bibfnamefont {H.}~\bibnamefont
  {St\"ocker}}, \bibinfo {author} {\bibfnamefont {M.~R.}\ \bibnamefont
  {Strayer}}, \ and\ \bibinfo {author} {\bibfnamefont {W.}~\bibnamefont
  {Greiner}},\ }\href {\doibase 10.1103/PhysRevLett.55.2786} {\bibfield
  {journal} {\bibinfo  {journal} {Phys. Rev. Lett.}\ }\textbf {\bibinfo
  {volume} {55}},\ \bibinfo {pages} {2786} (\bibinfo {year}
  {1985})}\BibitemShut {NoStop}%
\bibitem [{\citenamefont {Bai}\ \emph {et~al.}(1987)\citenamefont {Bai},
  \citenamefont {Cusson}, \citenamefont {Wu}, \citenamefont {Reinhard},
  \citenamefont {Stoecker}, \citenamefont {Greiner},\ and\ \citenamefont
  {Strayer}}]{Bai1987TDCDFT}%
  \BibitemOpen
  \bibfield  {author} {\bibinfo {author} {\bibfnamefont {J.~J.}\ \bibnamefont
  {Bai}}, \bibinfo {author} {\bibfnamefont {R.~Y.}\ \bibnamefont {Cusson}},
  \bibinfo {author} {\bibfnamefont {J.}~\bibnamefont {Wu}}, \bibinfo {author}
  {\bibfnamefont {P.~G.}\ \bibnamefont {Reinhard}}, \bibinfo {author}
  {\bibfnamefont {H.}~\bibnamefont {Stoecker}}, \bibinfo {author}
  {\bibfnamefont {W.}~\bibnamefont {Greiner}}, \ and\ \bibinfo {author}
  {\bibfnamefont {M.~R.}\ \bibnamefont {Strayer}},\ }\href {\doibase
  10.1007/BF01297581} {\bibfield  {journal} {\bibinfo  {journal} {Z. Phys. A}\
  }\textbf {\bibinfo {volume} {326}},\ \bibinfo {pages} {269} (\bibinfo {year}
  {1987})}\BibitemShut {NoStop}%
\bibitem [{\citenamefont {Vretenar}\ \emph {et~al.}(1993)\citenamefont
  {Vretenar}, \citenamefont {Berghammer},\ and\ \citenamefont
  {Ring}}]{Vretenar1993TDRMF}%
  \BibitemOpen
  \bibfield  {author} {\bibinfo {author} {\bibfnamefont {D.}~\bibnamefont
  {Vretenar}}, \bibinfo {author} {\bibfnamefont {H.}~\bibnamefont
  {Berghammer}}, \ and\ \bibinfo {author} {\bibfnamefont {P.}~\bibnamefont
  {Ring}},\ }\href {\doibase https://doi.org/10.1016/0370-2693(93)90776-E}
  {\bibfield  {journal} {\bibinfo  {journal} {Phys. Lett. B}\ }\textbf
  {\bibinfo {volume} {319}},\ \bibinfo {pages} {29} (\bibinfo {year}
  {1993})}\BibitemShut {NoStop}%
\bibitem [{\citenamefont {Vretenar}\ \emph {et~al.}(1995)\citenamefont
  {Vretenar}, \citenamefont {Berghammer},\ and\ \citenamefont
  {Ring}}]{VRETENAR1995TDRMF}%
  \BibitemOpen
  \bibfield  {author} {\bibinfo {author} {\bibfnamefont {D.}~\bibnamefont
  {Vretenar}}, \bibinfo {author} {\bibfnamefont {H.}~\bibnamefont
  {Berghammer}}, \ and\ \bibinfo {author} {\bibfnamefont {P.}~\bibnamefont
  {Ring}},\ }\href {\doibase https://doi.org/10.1016/0375-9474(94)00417-L}
  {\bibfield  {journal} {\bibinfo  {journal} {Nucl. Phys. A}\ }\textbf
  {\bibinfo {volume} {581}},\ \bibinfo {pages} {679} (\bibinfo {year}
  {1995})}\BibitemShut {NoStop}%
\bibitem [{\citenamefont {Tanimura}\ \emph {et~al.}(2015)\citenamefont
  {Tanimura}, \citenamefont {Hagino},\ and\ \citenamefont
  {Liang}}]{tanimura20153d}%
  \BibitemOpen
  \bibfield  {author} {\bibinfo {author} {\bibfnamefont {Y.}~\bibnamefont
  {Tanimura}}, \bibinfo {author} {\bibfnamefont {K.}~\bibnamefont {Hagino}}, \
  and\ \bibinfo {author} {\bibfnamefont {H.~Z.}\ \bibnamefont {Liang}},\ }\href
  {\doibase 10.1093/ptep/ptv083} {\bibfield  {journal} {\bibinfo  {journal}
  {Prog. Theor. Exp. Phys.}\ }\textbf {\bibinfo {volume} {2015}},\ \bibinfo
  {pages} {073D01} (\bibinfo {year} {2015})}\BibitemShut {NoStop}%
\bibitem [{\citenamefont {Ren}\ \emph {et~al.}(2017)\citenamefont {Ren},
  \citenamefont {Zhang},\ and\ \citenamefont {Meng}}]{REN2017Dirac3D}%
  \BibitemOpen
  \bibfield  {author} {\bibinfo {author} {\bibfnamefont {Z.~X.}\ \bibnamefont
  {Ren}}, \bibinfo {author} {\bibfnamefont {S.~Q.}\ \bibnamefont {Zhang}}, \
  and\ \bibinfo {author} {\bibfnamefont {J.}~\bibnamefont {Meng}},\ }\href
  {\doibase 10.1103/PhysRevC.95.024313} {\bibfield  {journal} {\bibinfo
  {journal} {Phys. Rev. C}\ }\textbf {\bibinfo {volume} {95}},\ \bibinfo
  {pages} {024313} (\bibinfo {year} {2017})}\BibitemShut {NoStop}%
\bibitem [{\citenamefont {Long}\ \emph {et~al.}(2004)\citenamefont {Long},
  \citenamefont {Meng}, \citenamefont {Van~Giai},\ and\ \citenamefont
  {Zhou}}]{Long2004PK1}%
  \BibitemOpen
  \bibfield  {author} {\bibinfo {author} {\bibfnamefont {W.}~\bibnamefont
  {Long}}, \bibinfo {author} {\bibfnamefont {J.}~\bibnamefont {Meng}}, \bibinfo
  {author} {\bibfnamefont {N.}~\bibnamefont {Van~Giai}}, \ and\ \bibinfo
  {author} {\bibfnamefont {S.-G.}\ \bibnamefont {Zhou}},\ }\href {\doibase
  10.1103/PhysRevC.69.034319} {\bibfield  {journal} {\bibinfo  {journal} {Phys.
  Rev. C}\ }\textbf {\bibinfo {volume} {69}},\ \bibinfo {pages} {034319}
  (\bibinfo {year} {2004})}\BibitemShut {NoStop}%
\bibitem [{\citenamefont {Lalazissis}\ \emph {et~al.}(2005)\citenamefont
  {Lalazissis}, \citenamefont {Nik\ifmmode \check{s}\else
  \v{s}\fi{}i\ifmmode~\acute{c}\else \'{c}\fi{}}, \citenamefont {Vretenar},\
  and\ \citenamefont {Ring}}]{lalazissis2005new}%
  \BibitemOpen
  \bibfield  {author} {\bibinfo {author} {\bibfnamefont {G.~A.}\ \bibnamefont
  {Lalazissis}}, \bibinfo {author} {\bibfnamefont {T.}~\bibnamefont
  {Nik\ifmmode \check{s}\else \v{s}\fi{}i\ifmmode~\acute{c}\else \'{c}\fi{}}},
  \bibinfo {author} {\bibfnamefont {D.}~\bibnamefont {Vretenar}}, \ and\
  \bibinfo {author} {\bibfnamefont {P.}~\bibnamefont {Ring}},\ }\href {\doibase
  10.1103/PhysRevC.71.024312} {\bibfield  {journal} {\bibinfo  {journal} {Phys.
  Rev. C}\ }\textbf {\bibinfo {volume} {71}},\ \bibinfo {pages} {024312}
  (\bibinfo {year} {2005})}\BibitemShut {NoStop}%
\bibitem [{\citenamefont {Nik\ifmmode \check{s}\else
  \v{s}\fi{}i\ifmmode~\acute{c}\else \'{c}\fi{}}\ \emph
  {et~al.}(2008)\citenamefont {Nik\ifmmode \check{s}\else
  \v{s}\fi{}i\ifmmode~\acute{c}\else \'{c}\fi{}}, \citenamefont {Vretenar},\
  and\ \citenamefont {Ring}}]{Niksic2008DDPC1}%
  \BibitemOpen
  \bibfield  {author} {\bibinfo {author} {\bibfnamefont {T.}~\bibnamefont
  {Nik\ifmmode \check{s}\else \v{s}\fi{}i\ifmmode~\acute{c}\else \'{c}\fi{}}},
  \bibinfo {author} {\bibfnamefont {D.}~\bibnamefont {Vretenar}}, \ and\
  \bibinfo {author} {\bibfnamefont {P.}~\bibnamefont {Ring}},\ }\href {\doibase
  10.1103/PhysRevC.78.034318} {\bibfield  {journal} {\bibinfo  {journal} {Phys.
  Rev. C}\ }\textbf {\bibinfo {volume} {78}},\ \bibinfo {pages} {034318}
  (\bibinfo {year} {2008})}\BibitemShut {NoStop}%
\bibitem [{\citenamefont {Zhao}\ \emph {et~al.}(2010)\citenamefont {Zhao},
  \citenamefont {Li}, \citenamefont {Yao},\ and\ \citenamefont
  {Meng}}]{ZhaoPC-PK1}%
  \BibitemOpen
  \bibfield  {author} {\bibinfo {author} {\bibfnamefont {P.~W.}\ \bibnamefont
  {Zhao}}, \bibinfo {author} {\bibfnamefont {Z.~P.}\ \bibnamefont {Li}},
  \bibinfo {author} {\bibfnamefont {J.~M.}\ \bibnamefont {Yao}}, \ and\
  \bibinfo {author} {\bibfnamefont {J.}~\bibnamefont {Meng}},\ }\href {\doibase
  10.1103/PhysRevC.82.054319} {\bibfield  {journal} {\bibinfo  {journal} {Phys.
  Rev. C}\ }\textbf {\bibinfo {volume} {82}},\ \bibinfo {pages} {054319}
  (\bibinfo {year} {2010})}\BibitemShut {NoStop}%
\bibitem [{\citenamefont {B\"urvenich}\ \emph {et~al.}(2002)\citenamefont
  {B\"urvenich}, \citenamefont {Madland}, \citenamefont {Maruhn},\ and\
  \citenamefont {Reinhard}}]{Burvenich2002PCF1}%
  \BibitemOpen
  \bibfield  {author} {\bibinfo {author} {\bibfnamefont {T.}~\bibnamefont
  {B\"urvenich}}, \bibinfo {author} {\bibfnamefont {D.~G.}\ \bibnamefont
  {Madland}}, \bibinfo {author} {\bibfnamefont {J.~A.}\ \bibnamefont {Maruhn}},
  \ and\ \bibinfo {author} {\bibfnamefont {P.-G.}\ \bibnamefont {Reinhard}},\
  }\href {\doibase 10.1103/PhysRevC.65.044308} {\bibfield  {journal} {\bibinfo
  {journal} {Phys. Rev. C}\ }\textbf {\bibinfo {volume} {65}},\ \bibinfo
  {pages} {044308} (\bibinfo {year} {2002})}\BibitemShut {NoStop}%
\bibitem [{\citenamefont {Maruhn}\ \emph {et~al.}(2014)\citenamefont {Maruhn},
  \citenamefont {Reinhard}, \citenamefont {Stevenson},\ and\ \citenamefont
  {Umar}}]{Maruhn2014CPC}%
  \BibitemOpen
  \bibfield  {author} {\bibinfo {author} {\bibfnamefont {J.~A.}\ \bibnamefont
  {Maruhn}}, \bibinfo {author} {\bibfnamefont {P.~G.}\ \bibnamefont
  {Reinhard}}, \bibinfo {author} {\bibfnamefont {P.~D.}\ \bibnamefont
  {Stevenson}}, \ and\ \bibinfo {author} {\bibfnamefont {A.~S.}\ \bibnamefont
  {Umar}},\ }\href {\doibase http://dx.doi.org/10.1016/j.cpc.2014.04.008}
  {\bibfield  {journal} {\bibinfo  {journal} {Comput. Phys. Commun.}\ }\textbf
  {\bibinfo {volume} {185}},\ \bibinfo {pages} {2195} (\bibinfo {year}
  {2014})}\BibitemShut {NoStop}%
\bibitem [{\citenamefont {Eastwood}\ and\ \citenamefont
  {Brownrigg}(1979)}]{eastwood1979remarks}%
  \BibitemOpen
  \bibfield  {author} {\bibinfo {author} {\bibfnamefont {J.}~\bibnamefont
  {Eastwood}}\ and\ \bibinfo {author} {\bibfnamefont {D.}~\bibnamefont
  {Brownrigg}},\ }\href {\doibase https://doi.org/10.1016/0021-9991(79)90139-6}
  {\bibfield  {journal} {\bibinfo  {journal} {J. Comput. Phys.}\ }\textbf
  {\bibinfo {volume} {32}},\ \bibinfo {pages} {24} (\bibinfo {year}
  {1979})}\BibitemShut {NoStop}%
\end{thebibliography}

%

\end{document}